\documentclass[sigconf,screen]{acmart}
\usepackage{soul}
\renewcommand{\hl}[1]{#1}
 
\usepackage{amssymb}
\usepackage{amsfonts}
\usepackage[mathscr]{eucal}
\usepackage{algorithmic}
\usepackage{units}
\usepackage{soul} % For strikethrough text
\usepackage{gensymb}
\usepackage{hyperref}

\usepackage[ruled]{algorithm2e} % For algorithms

\SetAlFnt{\small}
\SetAlCapFnt{\small}
\SetAlCapNameFnt{\small}
\SetAlCapHSkip{0pt}

\usepackage{graphicx,wrapfig}

\newcommand{\DelRev}[1]{} % Revision (deleted text)

\newcommand{\figurewidth}{\linewidth} %double column
\newcommand{\customtilde}{{\raise.17ex\hbox{$\scriptstyle\sim$}}}

\newcommand{\figref}[1]{Figure~\ref{#1}}

\newcommand{\set}[1]{\mathcal{#1}}
\AtBeginDocument{%
  \providecommand\BibTeX{{%
    \normalfont B\kern-0.5em{\scshape i\kern-0.25em b}\kern-0.8em\TeX}}}

\copyrightyear{2025}
\acmYear{2025}
\setcopyright{cc}
\setcctype{by}
\acmConference[UIST '25]{The 38th Annual ACM Symposium on User Interface Software and Technology}{September 28-October 1, 2025}{Busan, Republic of Korea}
\acmBooktitle{The 38th Annual ACM Symposium on User Interface Software and Technology (UIST '25), September 28-October 1, 2025, Busan, Republic of Korea}\acmDOI{10.1145/3746059.3747771}
\acmISBN{979-8-4007-2037-6/2025/09}

\begin{document}

\title{Sculptable Mesh Structures for Room-Scale Form-Finding}


\author{Jesse T. Gonzalez}
\email{jtgonzal@cs.cmu.edu}
\affiliation{%
  \institution{Carnegie Mellon University}
  \city{Pittsburgh}
  \country{USA}
}
\author{Yanzhen Zhang}
\email{18388402982@163.com}
\affiliation{%
  \institution{Carnegie Mellon University}
  \city{Pittsburgh}
  \country{USA}
}
\author{Dian Zhu}
\email{dianz@andrew.cmu.edu}
\affiliation{%
  \institution{Carnegie Mellon University}
  \city{Pittsburgh}
  \country{USA}
}
\author{Alice Yu}
\email{xinranyu@andrew.cmu.edu}
\affiliation{%
  \institution{Carnegie Mellon University}
  \city{Pittsburgh}
  \country{USA}
}
\author{Sapna Tayal}
\email{sapnat@andrew.cmu.edu}
\affiliation{%
  \institution{Carnegie Mellon University}
  \city{Pittsburgh}
  \country{USA}
}
\author{Nazm Furniturewala}
\email{nfurnitu@andrew.cmu.edu}
\affiliation{%
  \institution{Carnegie Mellon University}
  \city{Pittsburgh}
  \country{USA}
}
\author{Ziying Qi}
\email{ziyingq@andrew.cmu.edu}
\affiliation{%
  \institution{Carnegie Mellon University}
  \city{Pittsburgh}
  \country{USA}
}
\author{Somin Ella Moon}
\email{sominm@andrew.cmu.edu}
\affiliation{%
  \institution{Carnegie Mellon University}
  \city{Pittsburgh}
  \country{USA}
}
\author{Leyi Han}
\email{leyih@andrew.cmu.edu}
\affiliation{%
  \institution{Carnegie Mellon University}
  \city{Pittsburgh}
  \country{USA}
}
\author{Alexandra Ion}
\email{alexandraion@cmu.edu}
\affiliation{%
  \institution{Carnegie Mellon University}
  \city{Pittsburgh}
  \country{USA}
}
\author{Scott E. Hudson}
\email{scott.hudson@cs.cmu.edu}
\affiliation{%
  \institution{Carnegie Mellon University}
  \city{Pittsburgh}
  \country{USA}
}
%%
%% By default, the full list of authors will be used in the page headers. This command allows the author to define a more concise list of authors' names for this purpose.
\renewcommand{\shortauthors}{Lin, et al.}

\begin{abstract}

It can be hard to design a physical structure entirely within the confines of a computer monitor. To better capture the interplay between real-world objects and a designer’s work-in-progress, practitioners will often go through a sequence of low-fidelity prototypes (paper, clay, foam) before arriving at a form that satisfies both functional and aesthetic concerns. While necessary, this model-making process can be quite time-consuming, particularly at larger scales, and the resulting geometry can be difficult to translate into a CAD environment, where it will be further refined.

This paper introduces a user-adjustable, room-scale, "shape-aware" mesh structure for low-fidelity prototyping. A user physically manipulates the mesh by lengthening and shortening the edges, altering the overall curvature and sculpting coarse forms. The edges are equipped with resistive length sensors, and transmit their configuration to a central computer. The structure can later be reproduced in software, connecting this prototyping stage to the larger computational design pipeline.

\end{abstract}





\begin{teaserfigure}
  \includegraphics[width=\textwidth]{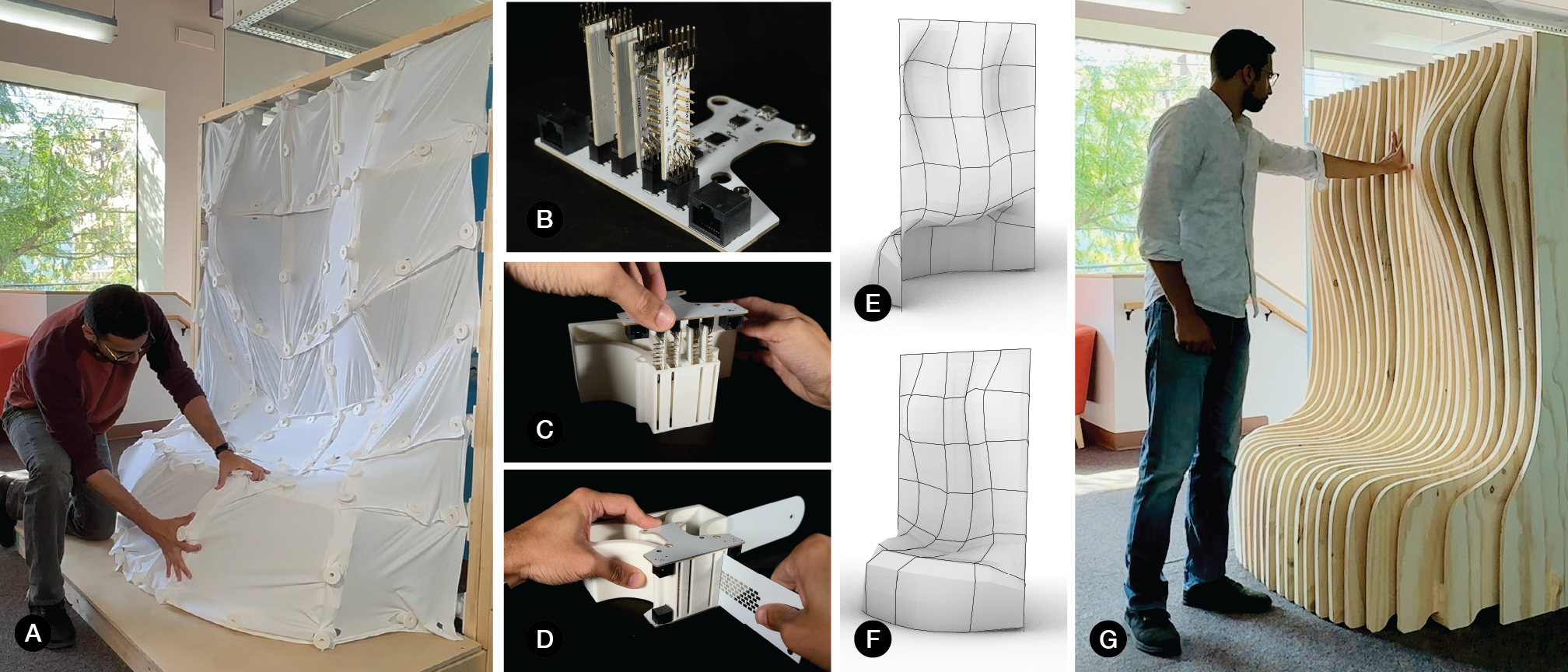}
  \caption{
    A large-scale mesh structure (a) composed of user-adjustable, flexible beams. A person can manipulate the structure by hand in order to sculpt the mesh into different forms. Each mesh edge is equipped with a length sensor (b-d), and the modules can transmit their configuration to an external computer, where the geometry can be reconstructed (e-f). The now-digitized geometry can serve as input to computer-controlled manufacturing processes, resulting in permanent, room-scale installations that have been \hl{"sculpted"} by non-CAD users (g). 
  }
  \label{fig:teaser}
\end{teaserfigure}

\maketitle

\section{Introduction}
\label{sec:introduction}

From architecture to engineering, computers allow us to construct physical forms that define our environments and captivate the eye. In the modern world, digital tools touch nearly every aspect of this process: from parametric modeling, to global file-sharing, to toolpath generation for computer-controlled fabrication. But the initial steps of a physical design — form-finding in particular — often benefit from a tactile approach that software cannot always provide. Designers need to assess how real-world objects will integrate with their creations with respect to both usability ( “Is the luggage rack too high?” “Do I have enough leg room?”) and aesthetic quality.  For this reason, practitioners will often make low-fidelity prototypes — many of them — out of foam, paper, or clay \cite{gulay2021understanding}. These mock-ups deliver practical insights in real-time, and in some cases (e.g. clay) allow the designer to make in-situ adjustments \cite{singh2006industrial}.

This hands-on step, however, is still quite siloed from the rest of the computational design pipeline. To bridge this gap, we introduce a room-scale, ``shape-aware'' mesh structure that can be hand-sculpted into a variety of forms — and can transmit its configuration to an accompanying software system in real-time. The structure itself is a flexible gridshell with adjustable-length members, capable of composing both single- and double-curved surfaces. It serves as a tangible interface to computer-based tools.

\begin{figure}[t]
    \centering
    \includegraphics[width=\figurewidth]{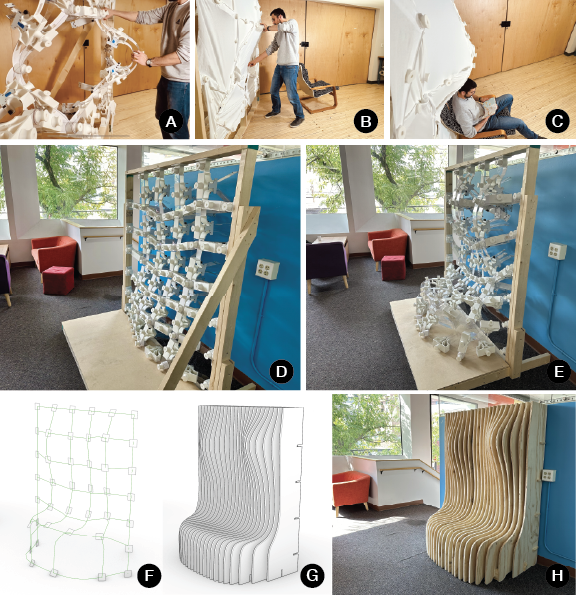}
    \caption{A user can shape our flexible gridshell structure by hand (a-c), as real-time sensor data is captured. For instance, a planar mesh (d) can be sculpted into an organic bench form (e). Since the members of this mesh are equipped with length sensors, the geometry can be reconstructed digitally (f,g) and fabricated (h).
    }
    \label{fig:overview}
\end{figure}

Interacting with our system is straightforward: a user walks up to the mesh and physically manipulates the edges and vertices, until they are satisfied with the resulting shape. Because the members are flexible (constructed from thin fiberglass panels), our mesh is well-suited for modeling curved, “organic” surfaces (\figref{fig:overview}). Resistive length sensors, embedded into each member, are used to capture this new geometry, smoothly connecting it to the full power of a computational modeling pipeline.

This type of direct interaction, outside of a computer, has a number of benefits. For user researchers, it allows for rapid iteration of tangible prototypes, combined with the state-saving advantages of a digital tool. For non-expert CAD users, it opens access to surface modeling, facilitating the co-design of architectural spaces.

Concretely, our contributions are:
\begin{enumerate}
    \item An extensible, room-scale mesh surface composed of user-adjustable, ``length-aware'' members.
    \item An energy-minimization routine for modeling bending-active structures composed of variable-length, flexible beams.
    \item An end-to-end demonstration of the workflow that our system enables, in the context of an architectural design and construction task.
\end{enumerate}

We particularly intend for our system to serve as a form-finding tool for architects and designers. A typical studio will have many tools for converting digital models into physical artifacts (laser cutters, 3D printers, etc); but the physical-to-digital workflow is less well-established. Our system adds another tool to the studio's arsenal: allowing designers to sculpt tangible forms, evaluate the prototypes on-location, and then use the sensed geometry to construct permanent installations. 

\hl{More broadly, HCI researchers have been increasingly interested in advancing tangible, shape-aware interfaces that provide direct manipulation capabilities and real-time feedback during the design process} \cite{leen2017strutmodeling, tahouni2020nurbsforms, weller2008posey}\hl{. We advance this space by introducing a system that combines multiple deformation capabilities (bending, length adjustment, and rotation) within a single unified interface.}\footnote{\hl{Implementation available at https://github.com/jtgonz/SculptableMesh}}

\section{Related Work}

Our approach bridges multiple research domains: design processes, tangible interfaces, and digital twins. To properly situate our work, we first examine the broader context of physical prototyping in design before addressing the specific gap our system targets.

\subsection{Physical Prototyping in Design Processes}

Design processes typically progress through phases of increasing fidelity and specificity \cite{sanders2014probes}. Within this progression, physical prototyping serves diverse purposes that vary by design stage and objective. Early explorations may use physical artifacts to inspire ideation, while subsequent stages might employ them to test functionality \cite{buchenau2000experience}. These low-fidelity iterations are important for understanding spatial relationships, ergonomics, and user interactions in ways that digital models alone cannot convey \cite{lim2008anatomy, buxton2010sketching}.  While architectural designers do often build smaller scale models of their designs, the complexities of human body-scale interactions with architecture \cite{pallasmaa_eyes_2012} can be much more directly investigated with full-scale models \cite{greenberg_value_2017}. 

Crucially, designers receive feedback from the properties and constraints of physical media. This material ``backtalk'' \cite{schon1992designing} often guides design decisions in ways difficult to replicate in purely digital environments. (For instance, when an architect manually bends a thin sheet of metal to explore a curved roof form, the physical resistance suggests natural resting states that inform the final design.) This type of embodied knowledge acquisition has been recognized as fundamental to architectural thinking \cite{binder2011design}.

We do not position our system as a replacement for all physical prototyping methods, many of which serve important purposes outside our scope. Instead, we focus specifically on the transition point between form-finding and digital capture — a persistent challenge in design workflows that has been identified as a significant bottleneck in the creative process \cite{evans2005rapid}.

\subsection{Digital Twins and Shape-Aware Materials}

Digital twins pair physical objects with virtual representations \cite{jones2020characterising} — in our case, embedded sensors capture the mesh's geometry for software reconstruction. Traditionally, digital twins are used in the latter stages of the product life cycle, to track usage patterns \cite{khajavi2019digital} or mechanical wear-and-tear \cite{glaessgen2012digital}. Our work instead leverages this concept earlier in the design process, bringing real-time digital capture into the initial prototyping phase.

In the HCI community, researchers have explored similar ideas through the development of "shape-aware" materials and structures \cite{wessely2018shape}. These are physical objects that detect user manipulation (i.e., extension, compression), and can stream this data to a connected computer. This can be accomplished with directly-embedded sensors \cite{piya2014proto}, or by leveraging proxy objects \cite{feick2020tangi} that work in conjunction with external cameras to track deformation \cite{sheng2006interface}.

These structures have proven valuable as creative input devices \cite{chien2015flexibend}. In puppetry applications \cite{weller2008posey}, for example, users can control digital characters by manipulating sensor-equipped "Tinkertoy-like" modules. Advances in piezoresistive materials \cite{bacher2016defsense} have further expanded these possibilities, enabling sensors to detect subtle physical interactions like squeezing or bending, and update digital models in real-time. Subtractive design approaches \cite{willis2010interactive} explore how digital models can update in response to physical cutting or carving. 

\hl{Particularly relevant to this work are devices designed to measure curves. ShapeTape} \cite{balakrishnan1999exploring, grossman2003interface} \hl{is an early example: a handheld rubber strip with fiber-optic bend sensors, which a user can manipulate in order to model and capture curves in 3D space. More recent methods have leveraged flexible circuit boards with onboard IMUs }\cite{dementyev2015sensortape}\hl{ (to capture twist), and capacitive sensing of shifting, offset electrodes }\cite{shahmiri2020sharc}\hl{.}

\hl{This has also led to the development of physical tools that also have a degree of spatial awareness. HandSCAPE} \cite{lee2000handscape} \hl{is an orientation-aware digital tape measure that captures both distance and direction, enabling users to digitize field measurements. Similarly, the SPATA tools} \cite{weichel2015spata} \hl{provide bidirectional transfer between physical and virtual measurements through actuated calipers and protractors.}

\hl{Further leveraging} this physical-digital link, researchers also have created tabletop construction kits that augment hands-on prototyping at the centimeter-scale. Systems like StrutModeling \cite{leen2017strutmodeling} and NurbsForms \cite{tahouni2020nurbsforms} allow users to assemble deformable physical components, which generate corresponding digital models.

Architectural design, however, often benefits from whole-body interaction and collaborative shaping. Our work scales up shape-aware principles to room-sized surfaces — addressing challenges in sensing and mechanical design to enable multiple users to physically sculpt and evaluate their creations at room-scale. This allows for the kind of intuitive, hands-on exploration that is crucial for processes such as participatory design.

\subsection{Room-Scale Physical Interfaces}

\hl{Within HCI, several approaches have emerged for quickly prototyping large physical forms. Using the handheld Protopiper extruder} \cite{agrawal2015protopiper}\hl{, for instance, a user can sketch physical wireframes of furniture-scale objects in 3D space. At larger scales, similar truss-like structures can be automatically deployed, from inflatable tubes that are either heat-sealed} \cite{swaminathan2019input} \hl{or tied in real-time} \cite{rambold2023airtied}\hl{. Researchers have also explored more modular approaches --- assemblies of electronically integrated structures using both rigid pipes }\cite{yu2022strawctures}\hl{ and polyhedral voxels} \cite{smith2025voxel}\hl{.}

\hl{Another direction focuses on reconfigurable architectural surfaces that can dynamically assume different functional forms }\cite{gonzalez2023constraint, je2021elevate, suzuki2020roomshift}\hl{. Though not explicitly designed to support direct manipulation by users, these systems demonstrate compelling methods of "rendering" low-fidelity forms in physical space. If coupled with an approachable user interface, such systems could be powerful tools for in-situ reshaping of environments.}

\subsection{Participatory Design in Architecture}
Participatory design actively involves stakeholders in the design process \cite{sanoff1999community}. Instead of designing \emph{for} users, architects using participatory techniques will design \emph{with} users, acknowledging that communities have unique insight into their own needs and daily patterns of life \cite{luck2018participatory}.

When spaces are designed this way, communities often feel greater ownership over the final result \cite{sanoff2002schools}. This collaborative approach can strengthen the connection between people and place. It has been shown to be an especially effective way of designing schools \cite{sanoff2002schools}, housing developments \cite{abrams2003byker}, and public areas \cite{sanoff2005community}.

The success of participatory design hinges on clear communication between architects and community participants. Here, physical prototypes serve as crucial tools. While architectural drawings and digital renderings can be challenging for non-professionals to interpret, physical prototypes provide an intuitive medium for spatial exploration. These tools bridge the expertise gap between architects and community members \cite{brandt2007tangible}—helping to focus discussions and test design ideas \cite{sanders2014probes}.

Particularly significant are "experiential prototypes"—physical models that allow people to directly interact with and experience spatial concepts \cite{stappers2009designing}. Full-scale mock-ups enable users to directly experience spatial relationships, helping them evaluate factors like ceiling heights, circulation paths, and furniture arrangements that might be unclear in traditional drawings \cite{sanoff1999community}.

By providing a tangible interface for spatial design while simultaneously capturing digital information, our system addresses a specific challenge in participatory design: how to involve community members meaningfully while maintaining a connection to the computational design pipeline that dominates modern architectural practice. This integration of physical exploration with digital design tools has the potential to enhance the participatory design process, making it more accessible and effective for everyone involved.

\section{Room-Scale Form-Finding with an Interactive Mesh}

Our large-scale, interactive gridshell is shown in \figref{fig:overview}. This is a mesh-like structure composed of bendable fiberglass members, whose lengths can be manually adjusted by a human operator. By manipulating these members, users can “sculpt” the mesh, transforming it from an initial configuration (i.e a flat plane) to a new one, analogous to how an artisan molds a block of clay. Though not load-bearing, the resulting shapes are stable, serving as a physical wireframe that designers can use to evaluate their forms. Onboard sensors detect the lengths of these wireframe edges, allowing a paired software system to capture and reconstruct the sculpted geometry for later use.

We envision an interaction workflow that follows the general pattern below:

\begin{enumerate}
    \item First, the user chooses an initial topology. In our examples, this is a quad mesh, bounded on three sides by a rigid frame. 
    \item Then, the user physically adjusts edges and vertices, while the accompanying software periodically saves the intermediate forms.
    \item Finally, the user leverages the digitized form to augment their design process. They may choose to review “snapshots” of their sculpted geometries, or import the digital mesh into a CAD tool for further processing.
\end{enumerate}

We detail this workflow in the following section, where we explore how a group of architects might use our tool to complete a small design and construction task.

\begin{figure}[t]
    \centering
    \includegraphics[width=\figurewidth]{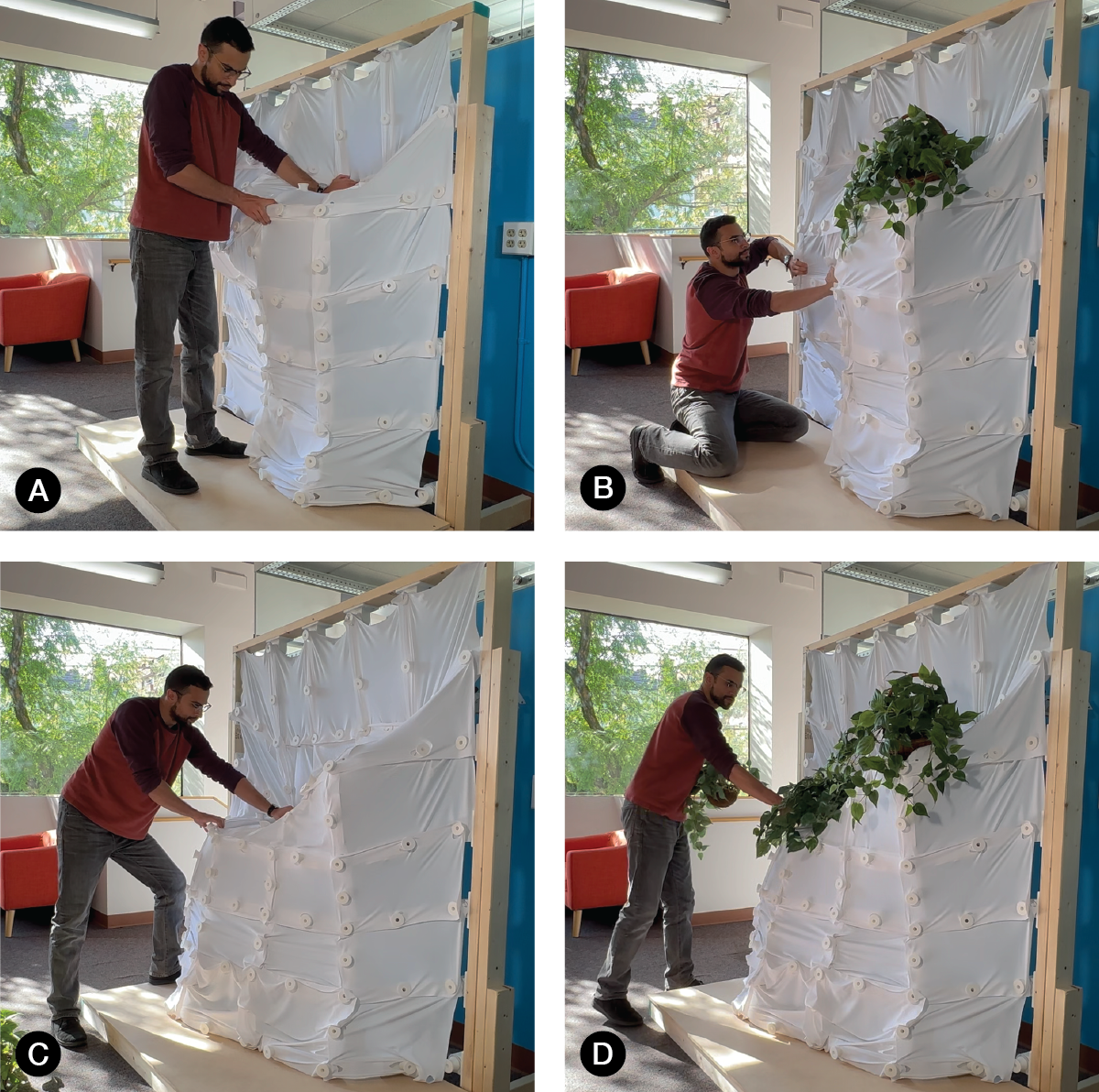}
    \caption{A person, inspired by the sunlight from a nearby window, sculpts a couple shelves to hold plants. \hl{The mesh holds its form and can support lightweight objects (such as the plants in b and d) without additional reinforcements.}
    }
    \label{fig:mesh-plants}
\end{figure}

\subsection{From Physical, to Digital, to Physical Again}

Consider the following example: a local library is undergoing a large renovation, and the architects involved would like to engage the community in the design process. They bring a few participants into the building, and direct them to an area with a large mesh wall. Here, the participants are instructed to shape the space into a structure that suits their needs, experimenting with different forms.

Looking towards a nearby window, one participant notes that this site receives a large amount of sunlight. They think it would be great spot to keep some plants, and begin shaping the mesh accordingly (\figref{fig:mesh-plants}). First, they pull out a section of the mesh and flatten it, turning it into a shelf. Then, inspired by the manner in which the plant leaves hang, they sculpt a second, lower shelf beside it. They choose to have this shelf taper off, so as not to obstruct the nearby stairway. Because this low-fidelity prototyping is taking place on-site, the participants can use features of surrounding environment (e.g. windows, sunlight, stairway) to stimulate and inform their design. 

\begin{figure}[t]
    \centering
    \includegraphics[width=\figurewidth]{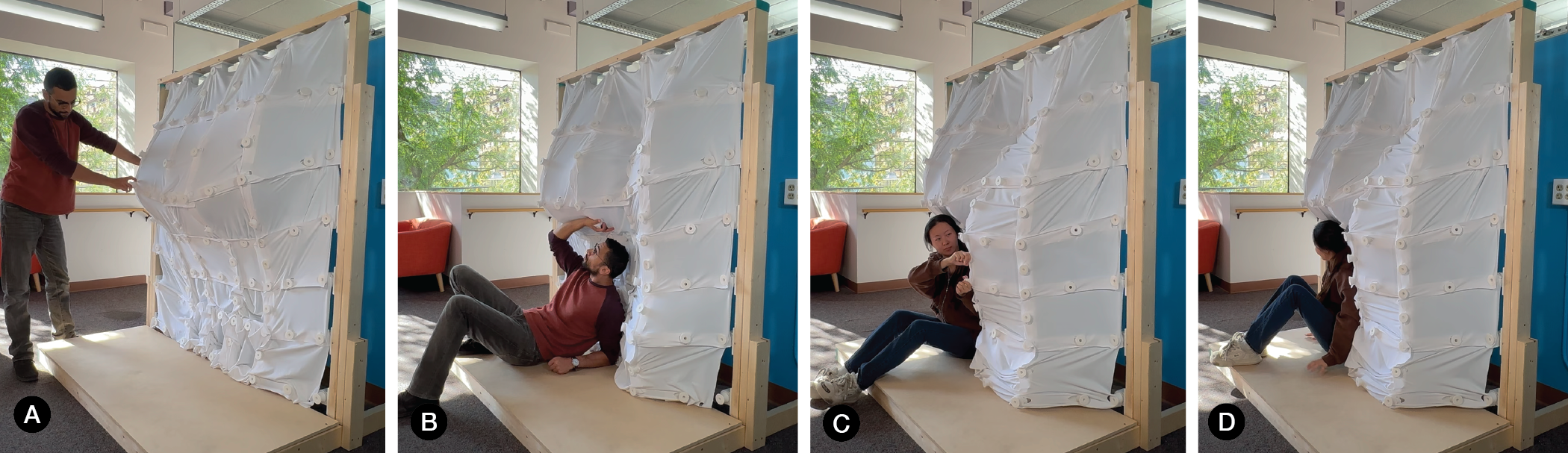}
    \caption{One person approaches the mesh wall and sculpts a child-sized alcove (a,b). A second person sits inside and modifies the alcove. \hl{She pulls the mesh around her so as to cover herself more fully (c), in an attempt to create a cozier environment (d).}
    }
    \label{fig:mesh-nook}
\end{figure}

Another participant decides to sculpt a small alcove, reasoning that children might enjoy hiding within the nook (\figref{fig:mesh-nook}). After an initial "draft", other participants sit inside, testing the fit. One participant feels that the nook is too shallow, so they drag a piece of the mesh outward, enveloping themselves for a cozier atmosphere (\figref{fig:mesh-nook}c). The architects take note of this and save the configuration digitally.

\begin{figure}[t]
    \centering
    \includegraphics[width=\figurewidth]{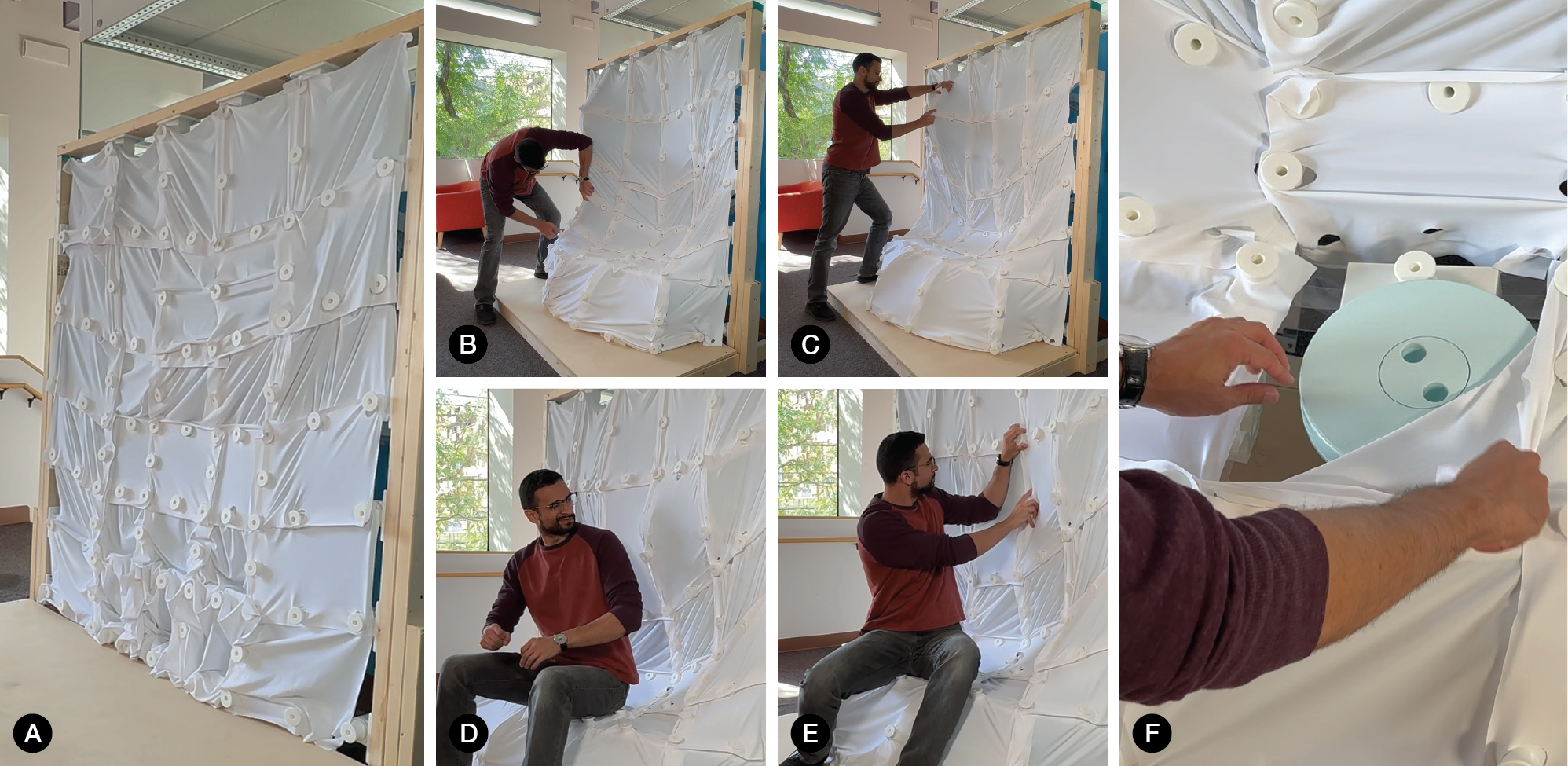}
    \caption{In its default state, our sculptable mesh surface is a flat plane (a). A person can pull on the edges and vertices to create forms such as a bench (b,c). They can use physical items, such a stool (f) to interact with their creations. In this example, a person notices that their sculpted bench is uncomfortable (d), and so they make a modification (e).
    }
    \label{fig:mesh-bench}
\end{figure}

The site is near the entrance of the library. Some participants point out that they often see people standing in this area, waiting for friends or family to arrive. They suggest that some additional seating could be useful here, and begin to sculpt a bench (\figref{fig:mesh-bench}). The participants also add a wave-like pattern to the wall, which they think compliments the shape of the seat.

The mesh itself is not meant to be load-bearing, so to test the bench, one participant grabs a nearby stool and places it inside of the mesh structure (\figref{fig:mesh-bench}f). When they sit down, they discover than the sculpted wave bulges out too far, and is uncomfortable. Though it appeared fine at first, physical testing made this design flaw apparent. The participant pushes the bulge inward, creating a cavity that they can lean back against (\figref{fig:mesh-bench}e). They note that this would be a nice spot to sit and read a book.  Ultimately, the participants conclude that — this being a library — the reading bench is the best choice for the space. 

\subsection{End-to-End Workflow}

With the community participants having finalized the bench design through physical sculpting and testing, the architects now proceed to transform this temporary prototype into a permanent installation. This process demonstrates how our system bridges the gap between intuitive physical prototyping and precise digital fabrication.

\begin{figure}[t]
    \centering
    \includegraphics[width=\figurewidth]{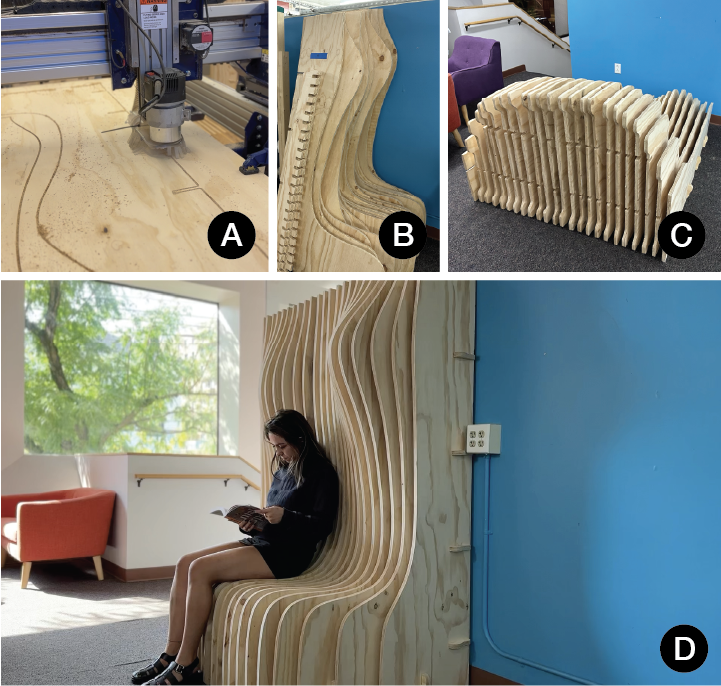}
    \caption{Using data from the length sensors, the sculpted bench is reconstructed in software, and can be connected to the rest of the digital manufacturing pipeline. In this example, the digital mesh is used to create CNC toolpaths (a), which then results in a permanent installation (d).
    }
    \label{fig:construct-bench}
\end{figure}

\begin{figure}[t]
    \centering
    \includegraphics[width=\figurewidth]{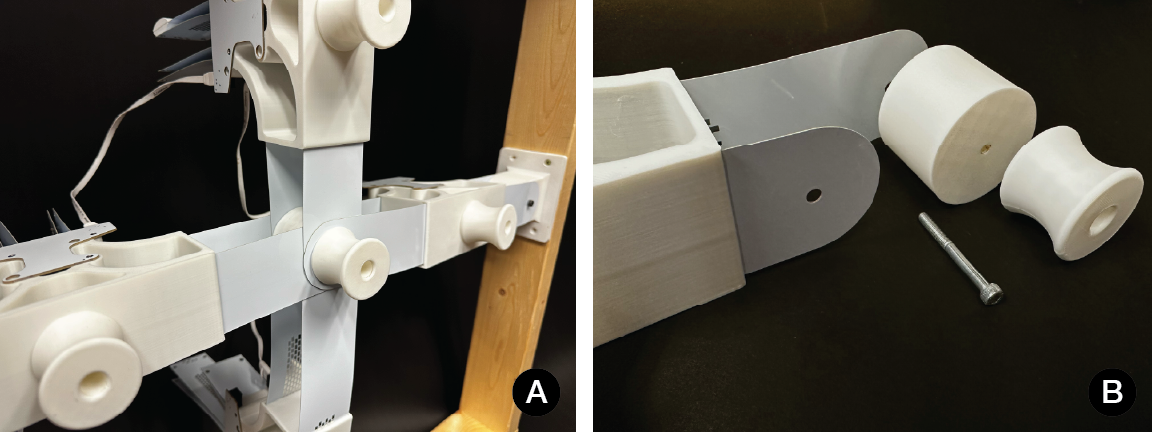}
    \caption{Modules can be joined together at the vertices (a) by using passive mechanical fasteners (b).
    }
    \label{fig:t-module-connect}
\end{figure}

Since each edge of the sculpted mesh is equipped with length sensors, the architects can capture the exact geometry through a software interface connected to the mesh. Length readings for all modules are continuously streamed to Rhino CAD software via this interface. Within seconds, the physical adjustments that the participants made (the height of the bench, the waves on the wall, the sculpted cavity) are preserved in a digital wireframe model.

The reconstruction algorithm (detailed in Section 6) processes this wireframe, first by applying energy minimization to establish the correct form, then by fitting cubic splines along each member and generating smooth Coons patches between them. This results in a continuous NURBS surface that captures the essential ergonomic features while creating a clean, fabrication-ready model (\figref{fig:overview}g).

For fabrication, the architects decide to slice the bench into panels that can be cut from standard plywood sheets --- a common fabrication strategy for complex curved forms. Using contour tools in Rhino, they generate evenly spaced vertical sections that, when stacked, will recreate the 3D form. These contours are arranged to maximize material efficiency and include alignment features to ensure accurate assembly.

The prepared profiles are exported as vector files and imported into CAM software for toolpath generation (\figref{fig:construct-bench}a). After the panels are cut, they are sanded, assembled according to their designed sequence, and press-fit together (\figref{fig:construct-bench}b,c).

The completed bench is installed in the exact location where the initial prototyping took place (\figref{fig:construct-bench}d). The library now has a permanent installation that was collaboratively sculpted by stakeholders in the community — a direct physical manifestation of their design input.

% \subsection{Further Configurations}

% klk Bring mesh to different sites. sdsdf.

% The overhang in \figref{fig:mesh-nook} may be 
%  \figref{fig:overhang}

% <Here we detail other envisioned uses,  \figref{fig:overhang}.>

% \begin{figure}[h]
%     \centering
%     \includegraphics[width=2.5in]{figures/renders@2x.png}
%     \caption{Envisioned uses. Top: An exhibit designer working with difficult-to-digitize items, such as plants, can arrange them directly on a physical display table. Bottom: Children can help design a play space without needing to understand scale models or virtual renderings.
%     }
%     \label{fig:renders}
% \end{figure}

\begin{figure}[t]
    \centering
    \includegraphics[width=\figurewidth]{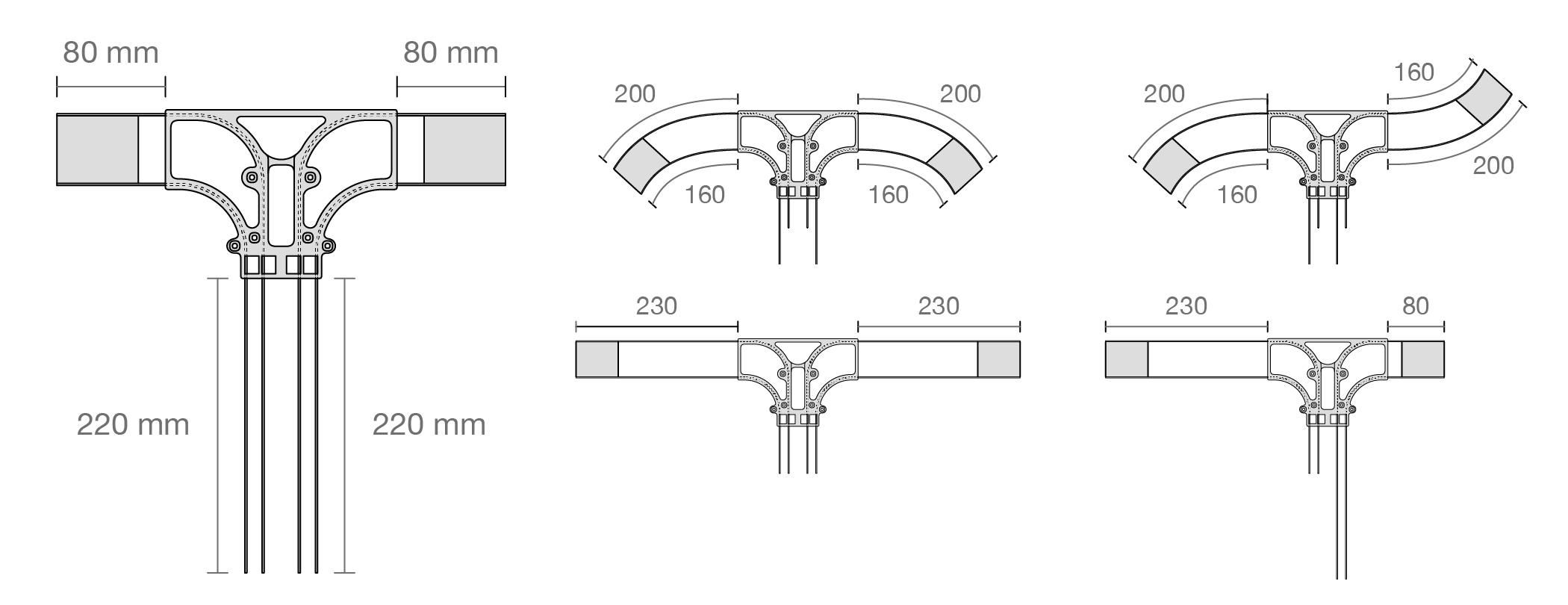}
    \caption{The "T-Module" is the unit cell of our sculptable mesh structure, acting as a variable-length edge. It consists of flexible bands routed through a rigid guide. When fully contracted, the module measures 22 cm from vertex to vertex. When fully expanded, it reaches 56 cm. Excess material is directed towards the interior of the mesh, away from the user.
    }
    \label{fig:t-module-diagram}
\end{figure}

\begin{figure}[t]
    \centering
    \includegraphics[width=\figurewidth]{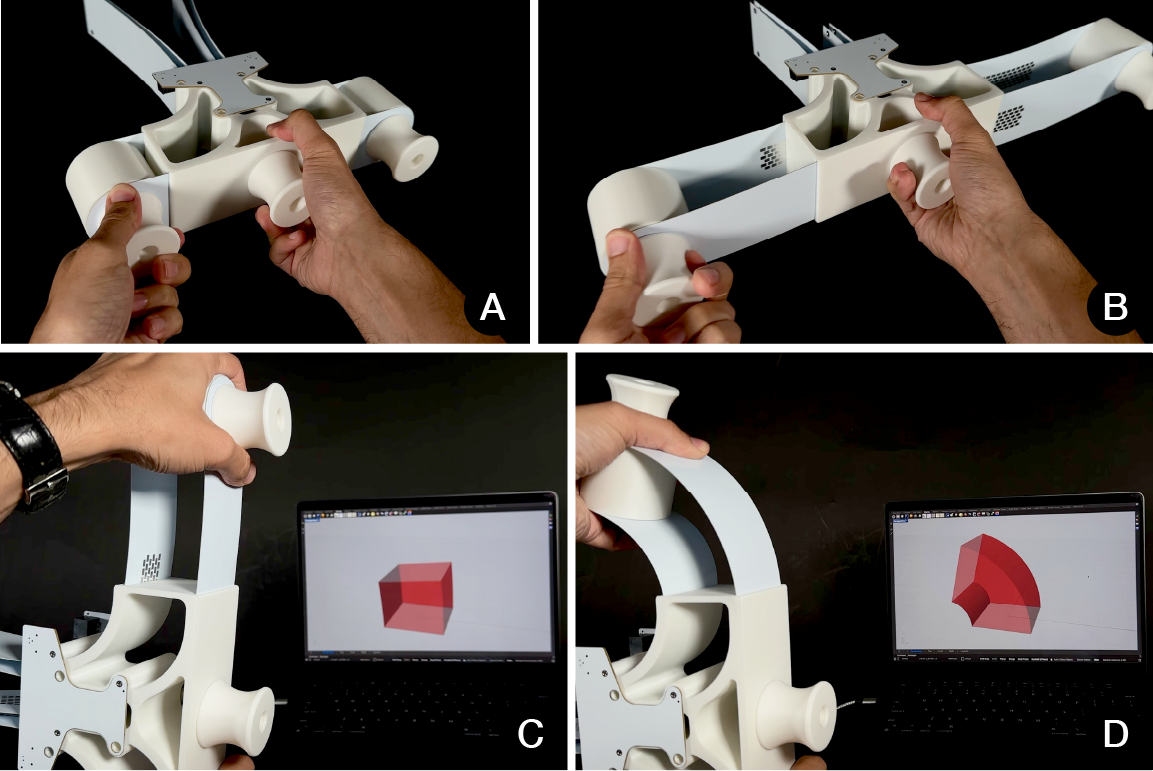}
    \caption{A user can alter the length and curvature of the T-Module by pulling on its handles (a,b). Each side of the module can be adjusted individually. Embedded length sensors capture the module geometry (c,d). By composing many of these members together, we can create sculptable forms that have an inherent digital twin.
    }
    \label{fig:basic-operation}
\end{figure}

\section{Overview and Mechanical Design}

The T-shaped structure shown in \figref{fig:t-module-diagram} is the “unit cell” of our sculptable mesh, and we refer to it as a "T-Module". It is an adjustable-length beam, made of four flexible FR4 \hl{(fiberglass)} bands which can slide freely through a plastic guide. A user can pull on these bands to expand a particular edge (\figref{fig:basic-operation}b), or slide them inward to shrink the edge, tucking the excess material away (\figref{fig:basic-operation}a).

\hl{These modules can be linked together to form a mesh, as shown in Figure 7. Mechanically, the T-Modules are connected with simple fasteners (plastic cylinders, with a metal bolt as an axle). Electrically, they are networked via flexible patch cables in a winged-edge configuration (see Section 5).}

Individually, the \hl{FR4 bands of the T-Module} are flexible enough to pass through the \hl{plastic} guide without much resistance, but together (in a “two-tier” configuration), they form a prismatic structure with enough rigidity to keep the parent mesh from collapsing. Note that for the curved forms in \figref{fig:t-module-diagram}, there is a length differential between the inner and outer bands, which allows us to detect the direction of curvature (\figref{fig:basic-operation}c,d).

The four FR4 strips are circuit boards\footnote{Fabricated externally, via JLCPCB.}, 0.8~mm thick. By adding these boards, we transform this passive module into a “length-aware”  mesh edge. Electrically, each circuit operates much like a slide potentiometer. The flexible boards are patterned with a resistive code, which, when read by a scanning head at the “neck” of the module, can be mapped to a physical displacement. A microcontroller, connected to this scanning head, records this length data and transmits it to an external computer. (This is done via neighbor-to-neighbor communication within the mesh — the message is passed from module to module until it reaches a pre-selected relay unit). The strips contain no onboard power source — instead, the scanning head sends a small current through the strip's resistor ladder, and takes a reading.

The modules are joined mechanically through passive vertices — plastic cylinders with a metal axle. They are joined electrically through flexible, Cat6 patch cables, in a winged edge configuration. Any individual module can serve as a relay device, collecting length data from the parent mesh and transmitting it to our software. (In practice, we choose a module near the outer perimeter to perform this function).

\begin{figure}[t]
    \centering
    \includegraphics[width=\figurewidth]{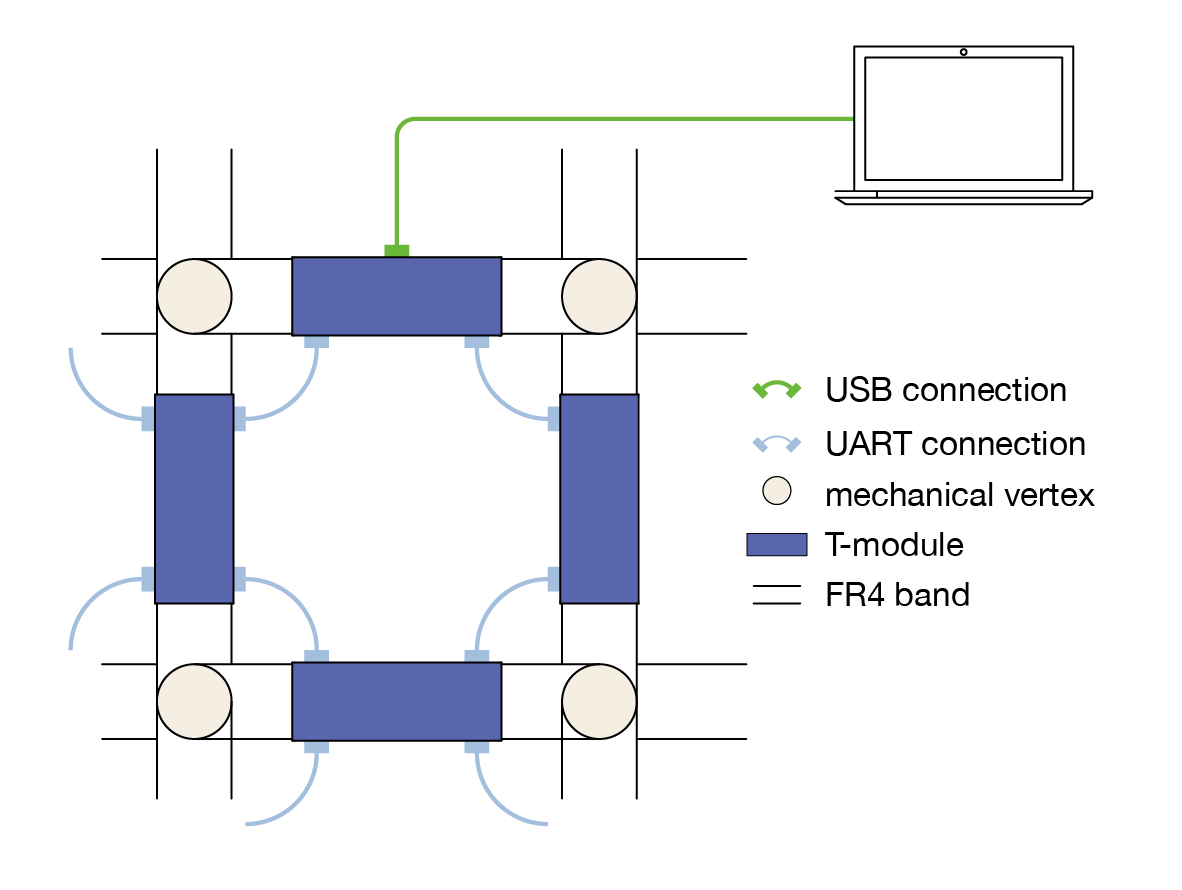}
    \caption{ Connection scheme for our sculptable mesh surface. T-Modules are joined to their neighbors using Cat6 cables, and exchange messages via a UART peripheral. One T-Module is connected to a computer, where is transmits the mesh edge lengths to our reconstruction software.
    }
    \label{fig:connections}
\end{figure}

\subsection{Curvature and Connection Scheme}

Our mesh structures mimic a biaxial weave (\figref{fig:connections}), with four edges meeting at every interior vertex. Compared to other patterns, this configuration gives us a good balance between user manipulability and structural stability. We found that denser structures, such as a valence-6 triangular mesh, were stiff near the vertices and difficult to physically adjust. Sparser structures, such as a valence-3 hexagonal mesh, too often fell into configurations that were underconstrained. A valence-4 configuration gave us a good balance between these alternatives.

\begin{figure}[t]
    \centering
    \includegraphics[width=\figurewidth]{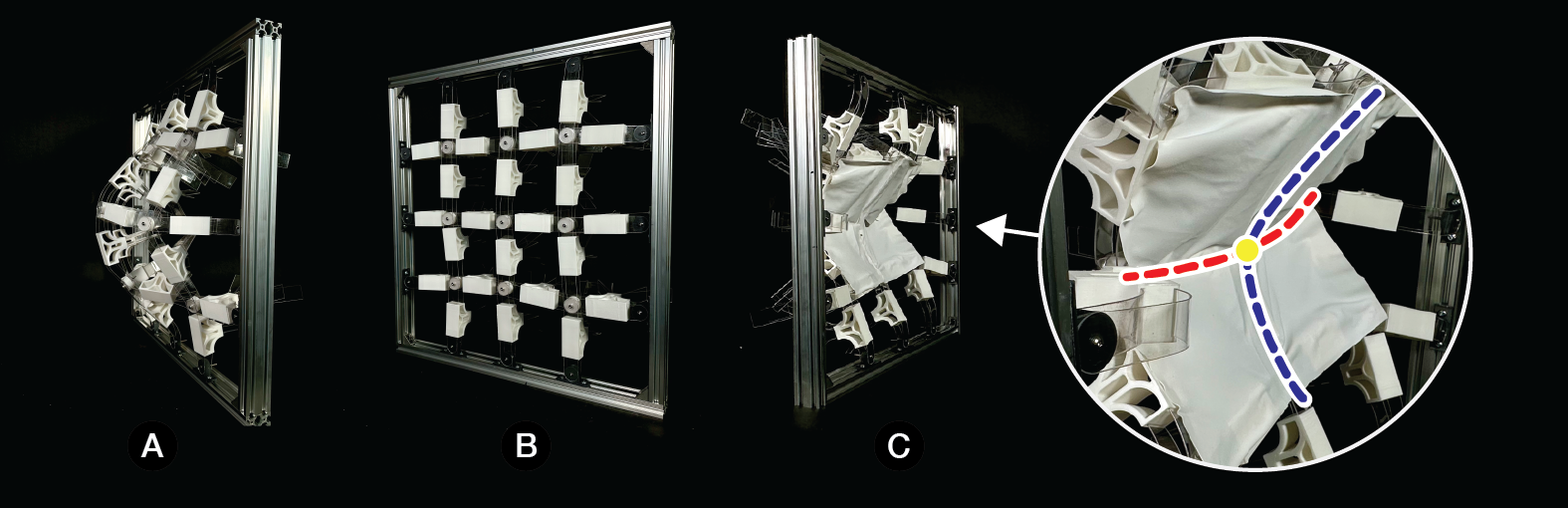}
    \caption{By expanding, contracting, and bending modules, our mesh can transition from a state of zero curvature (b) to a state of either positive curvature (a) or negative curvature (c). We have attached a piece of fabric to the modules in (c) to make the saddle point more apparent.
    }
    \label{fig:curvature}
\end{figure}

Our mesh structure is able to transition smoothly between single and double curvature (\figref{fig:curvature}). By adjusting the length of the flexible members, the surface can "stretch" in ways that would normally cause kinks or distortions in conventional materials. This allows us to adapt to both \hl{single-curved and double-curved surfaces}, without compromising the continuity of the form.

\hl{Though quad meshes are theoretically susceptible to shearing, we did not experience noticeable issues in our prototypes. This assumes that the boundaries of the mesh are anchored to fixed points, which provides sufficient constraint to prevent unwanted deformations.}

\subsection{Materials and Fabrication}

At the core of each module is a 3D-printed, PLA plastic guide. It contains four curved slots, each of which constrain a thin strip of FR4 at a bend radius of 50 mm. This radius was chosen to mitigate plastic deformation — at smaller radii, this deformation becomes noticeable by eye, and introduces additional stresses into the system (making it difficult to model in software). The strips themselves are 60 mm wide and 0.8 mm thick. The inner and outer strips are spaced 50 mm apart.

An early engineering challenge was optimizing the stiffness of the T-Modules. Overly stiff beams limit our ability to form curved surfaces; but if the beams are too flexible, the sculpted mesh won't hold its shape. Stiffness also impacts the extension ratio of the modules — as the strips pass through the rigid guide, stiffer strips must do so at a larger bend radius. A larger bend radius requires a longer \hl{plastic guide, which ultimately reduces the overall extension ratio.}

The "two-tier" prismatic structure \hl{helps solve} this problem: \hl{instead of directly optimizing the flexibility of the beam, we can optimize the distance between the outer and inner strips.} This allows us to experiment with different stiffness levels while using strips of a constant thickness \hl{--- a larger distance between strips corresponds to a stiffer beam.}

Additionally, using two thin bands instead of a single thick band permits a smaller bend radius at the rigid guide, which in turn increases the dynamic range of the adjustable edge.

% Stiffness also impacts the extension ratio of the modules — as the strips pass through the rigid guide, stiffer strips must do so at a larger bend radius. A larger bend radius requires a longer guide, which ultimately reduces the overall extension ratio. Our early prototypes used only a single pair of PETG strips (with no inner band), making it difficult to balance the trade-off between shape retention and extension ratio simply by adjusting the strip thickness.

% Before switching to FR4, we used 1.5 mm thick PETG strips as the primary element of this two-tier structure. To incorporate length sensing, flexible PCBs were adhered to the PETG strips using a double-sided adhesive tape (\figref{fig:interior}b). While PETG is less brittle than FR4, it is more prone to plastic deformation and proved to be a more expensive option overall due to the added cost of the flexible PCBs.

For ease of manipulation, we attach plastic knobs to the both the T-modules and mechanical vertices. Once a user adjusts the length of the beam, static friction keeps it in place. \hl{(We experimented with mechanical locking mechanisms for the beams, but found friction sufficient for furniture-scale modeling.)}

\subsection{Topology and Anchors}

In our current prototype, the user must provide a starting topology — in the form of an adjacency matrix — before we can reconstruct the mesh. The user also specifies which (if any) vertices should act as “anchors”, along with their starting positions. For instance, in the mesh shown in Section 2, the user must specify that the 16 vertices along the top and sides of the frame are anchor points. In the current prototype, the user supplies this information via a C\# script. In future versions it should not be hard to do this with a simple interface.

% The second benefit of the double-band structure is related to digital reconstruction — by comparing the relative length of the bands, we can determine the direction of curvature. We can then use this information to bias mesh vertices in a particular direction.

 % The bands on ola version uses FR4 strips, . earlier implementation used PETG strips, 1.5mm thick.
\begin{figure}[t]
    \centering
    \includegraphics[width=\figurewidth]{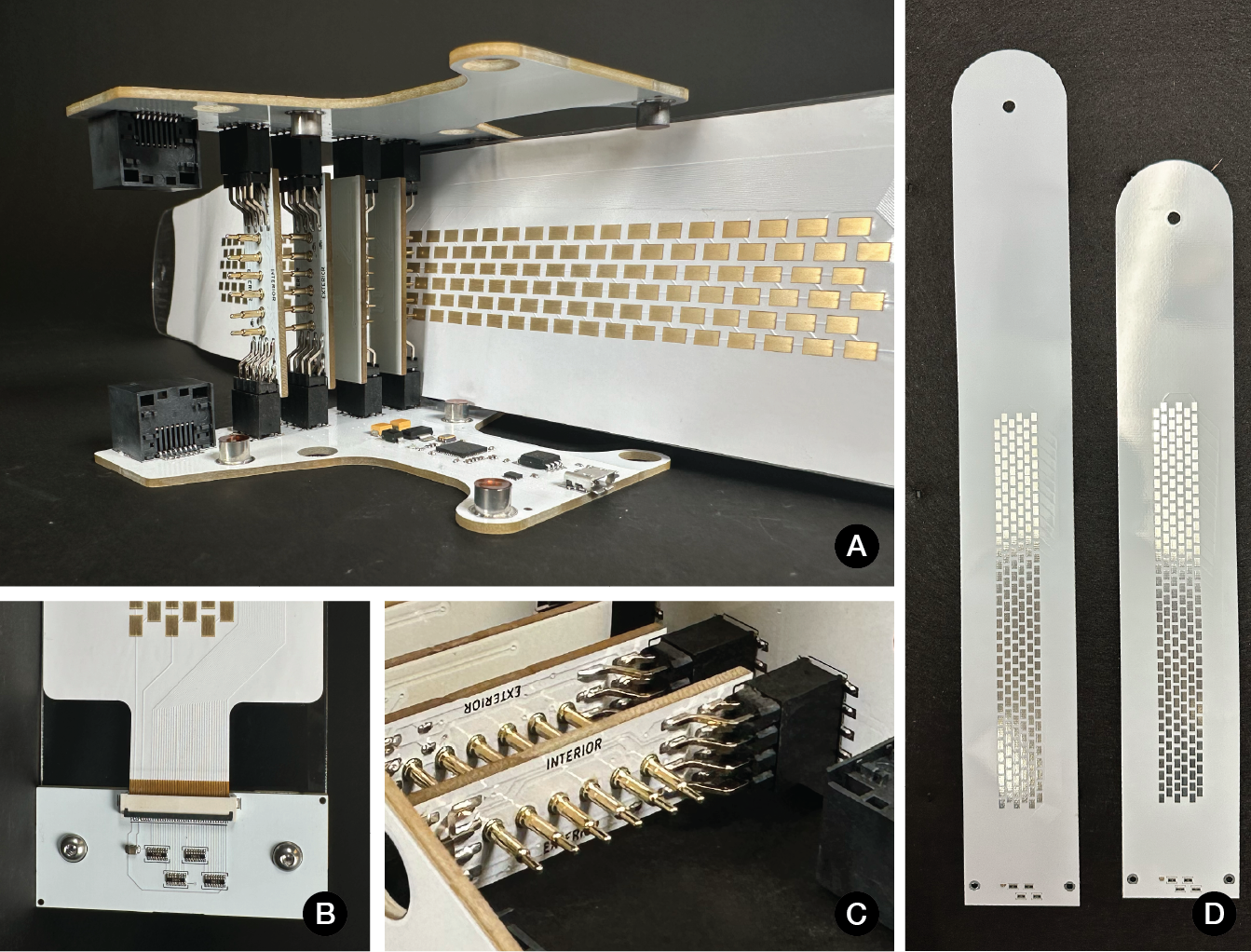}
    \caption{Interior of T-Module (a), which uses \hl{flexible FR4 strips (d)} as length sensors. The resistive code is read by a scanning head (c). Earlier prototypes used flexible printed circuit boards, attached to a rigid resistor bank (b).
    }
    \label{fig:interior}
\end{figure}

\section{Electronics and Firmware}

Every active module in our mesh structure is equipped with a small microcontroller\footnote{RP2040, by Raspberry Pi Ltd.}, which allows for local length sensing and neighbor-to-neighbor communication. Adjacent edges are connected by flexible cables, which transmit both power and data, allowing us to interface with the system via a single external connection.

\figref{fig:interior} shows an interior view of the module. Within, there are four components of interest:

\begin{enumerate}
    \item A faceplate, housing the microcontroller, sensing circuitry, a USB interface, and two power/communication ports.
    \item A backplate, providing two additional power/communication ports.
    \item  Two pairs of flexible circuits with a resistive ladder, each on an 0.8mm thick FR4 panel.
    \item  Two pairs of scanning heads, which read a resistive code from the flexible strips. These also facilitate power and data transmission between the faceplate and backplate.
\end{enumerate}

Note that while only one pair of strips (and scanning heads) are required for measuring the module length, an additional pair helps us determine the direction of curvature.

\subsection{Length Sensing}

When our software system first connects to the physical mesh, the length of each edge is unknown. This situation is similar to many machines found in an architecture studio, such as laser cutters or 3D printers, which address the issue by performing a homing sequence upon startup. However, unlike those machines, we cannot expect the user to manually "zero" each variable member before beginning the sculpting process — this would be far too tedious. As a result, absolute positioning becomes a critical requirement.

We need to sense changes in length along each \hl{strip}, detecting variations of up to 200 mm. Commercial sensors that can achieve this over such a range are either difficult to find or prohibitively expensive.

\hl{To meet this need, we implement a resistive ladder on each FR4 strip. The ladder itself is a chain of discrete 100-ohm resistors}\footnote{\hl{We found that this gave us a good balance between noise resistance and current draw.}}\hl{, connected in series at the bottom edge of the strip. Each connection point between resistors is then routed via a copper trace to an exposed pad positioned along the length of the strip.}

\hl{A scanning head with spring-loaded probes maintains contact with these copper pads. It operates much much like a slide potentiometer, measuring the voltage at whichever copper pad it currently contacts (Figure 14a). Since each pad corresponds to a specific position along the strip, the measured voltage directly indicates the module's length. This offers an inexpensive method for absolute length sensing, as the resistor ladder can be patterned on a flexible substrate (e.g., thin FR4 or polyimide) over significant distances.}

However, using only a single probe can cause issues when the scanning head crosses between pads (Figure 14b), as the probe is left floating. This ambiguity may be acceptable for incremental sensors, but it is insufficient for applications requiring absolute positioning, since the probe may start in an undefined state.

\begin{figure}[t]
    \centering
    \includegraphics[width=\figurewidth]{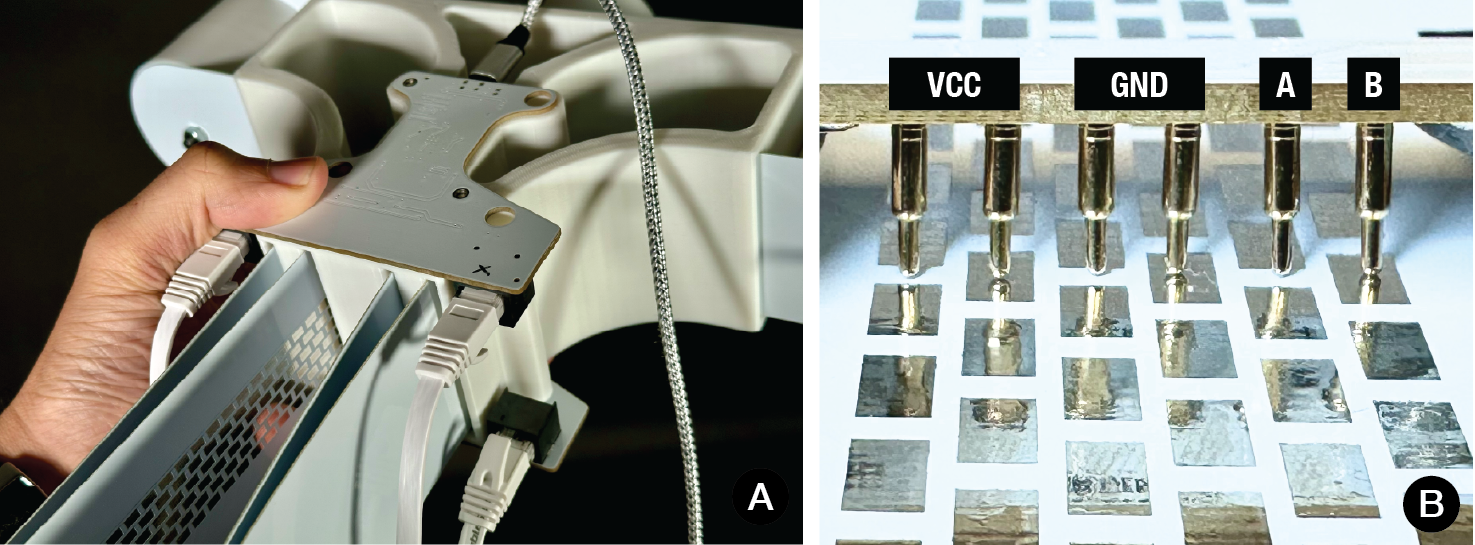}
    \caption{We detect module curvature by sensing the length of resistive strips (b). Having two probes on the scanning head allows us to avoid dead zones on the resistive ladder (see Figure 13a).
    }
    \label{fig:length-sense-photo}
\end{figure}

\begin{figure}[t]
    \centering
    \includegraphics[width=\figurewidth]{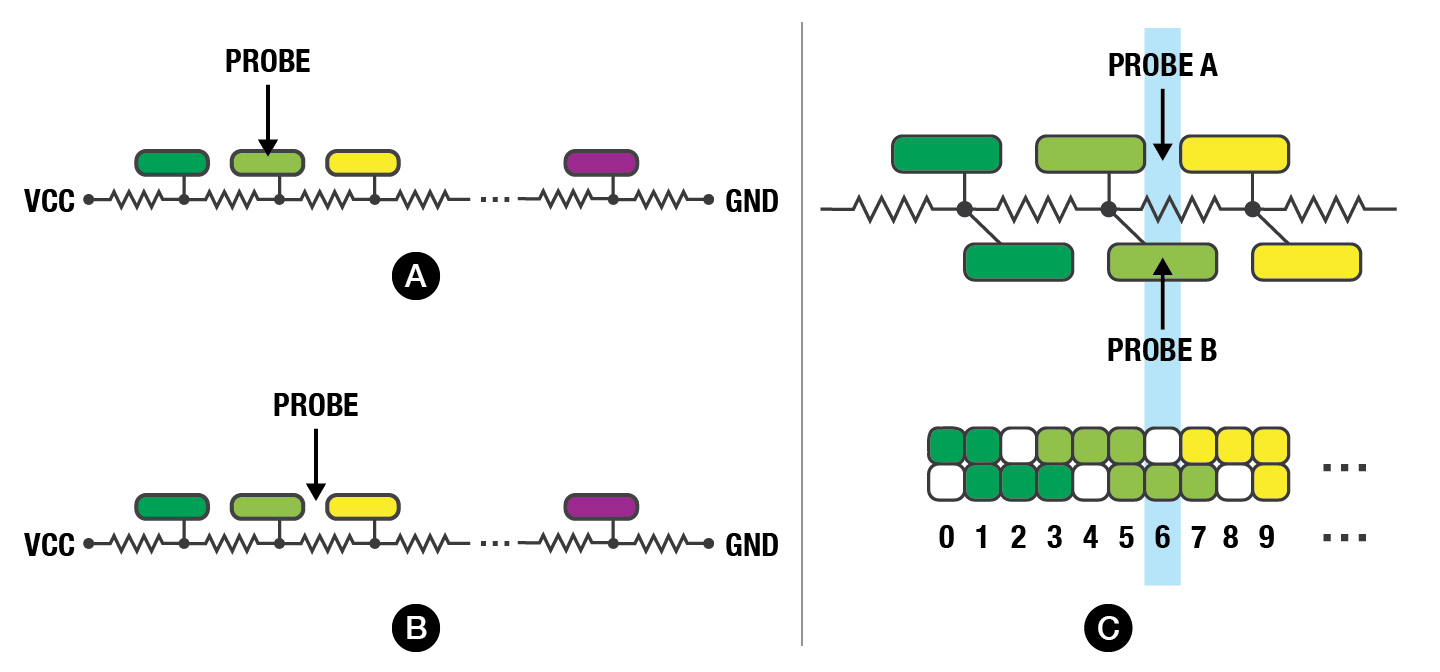}
    \caption{On the left (a,b) we show a standard resistive ladder; on the right (c) we show our configuration. Each node on our resistive ladder is connected to two copper pads, which are offset. This configuration ensures that there are no “dead zones” on the strip — areas between pads where the pins of the scanning head are not in contact with any part of the resistive ladder.
    }
    \label{fig:length-sense-diagram}
\end{figure}

We resolve this by introducing a second set of conductive pads, slightly offset from the first (\figref{fig:length-sense-diagram}b). In this staggered configuration, at least one probe is always in contact with a pad. This has the added benefit of increasing our resolution by a factor of four. With pad lengths of 6 mm, our resolution improves to 2 mm across a 200 mm active area.

\subsection{Curvature Detection}

With sensors on only the outer strips, we can detect edge length, but not the direction of curvature. For some scenarios, this is sufficient (for instance, if we know that the mesh is against a wall, we can assume that most edges will bend outward when lengthened, and then bias the simluation accordingly). However, in more underconstrained scenarios, we need to use sensors on both the top and bottom strips — which our system does indeed support.

With two probes per strip, we require a total of eight ADC channels to sense the length on all four strips. This actually exceeds the amount available on the RP2040 (which is limited to four), so we use an analog switch (Vishay DG2788A) to de-multiplex the inputs. By comparing the inner and outer strip lengths, we can determine which way the module bends.

\subsection{Power and Communication}

On the underside of each T-Module, there are four physical ports (\figref{fig:interior}a) for transmitting power and data between adjacent units. These connections are made through flexible Cat6 cables — although we use UART, rather than Ethernet, to communicate. Each unit passes messages directly to its neighbors, which allows us to easily extend and reconfigure the mesh network.

The RP2040 only features two hardware UARTs, but up to four additional interfaces can be implemented by using the onboard PIO state machines. We leverage all of them, and create four full duplex UART peripherals, so that each T-Module can communicate with its immediate neighbors. These four ports are sufficient, since no module will have more than four neighbors (as we explain in the following section).

Power can be injected at any point in the structure, \hl{or at multiple points}. This approach helps distribute the load \hl{in larger configurations}, reducing the current required through any individual cable and minimizing potential voltage drops across the system.

\subsection{Mesh Representation and Traversal}

\begin{figure}[h]
    \centering
    \includegraphics[width=\figurewidth]{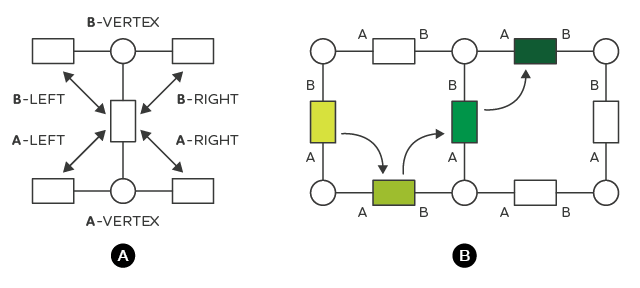}
    \caption{Our mesh can be represented by a winged edge data structure (a). Modules can send messages to each other by passing data from neighbor to neighbor (b). Along this path, modules construct a return address, so the receiver module can send a message back to the sender .
    }
    \label{fig:comms}
\end{figure}

In our implementation, neighboring modules are defined as mesh edges that (1) share a vertex and (2) belong to the same planar face. This definition allows us to efficiently represent our mesh using a winged-edge data structure \cite{baumgart1975polyhedron}. In this structure, each edge stores references not only to its two endpoints, \hl{but also to the two adjacent faces and the four edges that are incident to them} (\figref{fig:comms}a).

In contrast to other mesh representations (e.g. face-vertex or vertex-vertex), this representation provides several advantages for our system. First, it facilitates traversal of the mesh using only information encoded in the edges — important because the edges are the only active elements, capable of sending and receiving messages. Second, this structure supports efficient updates to the mesh topology, which is crucial as edges are adjusted during the sculpting process, enabling real-time interaction with the physical mesh.

One module sends a message to another by attaching an address — a list of hops — to a data packet and passing it to a neighboring module. Each module along the path tracks the number of hops and forwards the message to the next neighbor specified in the address list (\figref{fig:comms}b). As the message passes through, each module also notes the port through which it received the message and appends this information to the packet, effectively creating a return address. When the target module receives the message, it can reply to the original sender by following the return address generated during the message's journey.

\subsection{Physical-to-Digital Link}

To ensure smooth interaction between the physical mesh and its digital twin, we must establish a direct connection between the two systems. This process involves linking each physical module to the corresponding edge in the software model, allowing real-time communication and synchronization between the virtual and physical representations of the structure.

First, the user defines the mesh topology in software using a C\# script. Next, we need to align the physical mesh with the software-defined topology, meaning each physical module must be mapped to a corresponding edge in the software

For all of our applications, we choose one module to serve as the “relay unit”, and connect it to a central computer via USB. With this module as the root, we first perform a breadth-first traversal of the mesh, obtaining the IDs and addresses (routes relative to the root) of all connected units. After this initialization, the central computer begins polling each module, requesting sensor data. Each polled module responds by sending its ID and length data to the relay unit.

With the physical and digital meshes now synchronized, the user can manipulate the length of each edge in the physical structure, and these changes will be reflected within the digital model.
\section{Digital Reconstruction}

Once the lengths of each mesh member are known, we can reconstruct the overall geometry to create a digital twin of the physical structure. We approach this as a dynamic relaxation problem, using the Kangaroo physics engine \cite{piker2013kangaroo} within Rhino 8\footnote{Rhino 8 CAD software available at https://www.rhino3d.com/} to calculate and minimize the residual energy in our mesh structure.

For the mesh wall demonstrated in Section 2 (consisting of 54 T-Modules), our simulation creates approximately 300 particles. These particles correspond to physical points on the sculpted structure, and their positions are optimized through energy minimization.

At each iteration of our solver, a weighted sum of energies acting on each particle is calculated, a correction vector (pointing towards the lowest energy state) is constructed, and the particles are moved accordingly. This process continues until the system reaches a stable configuration that best matches the sensed physical state.

\subsection{Complete Reconstruction Pipeline}

Our digital reconstruction follows a multi-stage process:

\begin{enumerate}
    \item First, we initialize the mesh topology with known anchor positions (e.g. the wooden frame in which the physical mesh is suspended).
    \item We then perform a coarse solving pass using only length constraints between vertices, establishing the basic shape.
    \item Once this coarse mesh is established, we interpolate intermediate points along each member to represent the full T-Module geometry.
    \item Next, we add additional energy terms to refine the curvature and mechanical behavior of the mesh.
    \item Finally, we iteratively solve until the mesh represents a physically plausible configuration that satisfies our measured constraints.
\end{enumerate}

This approach allows us to capture the general physical form, preserving important distances relevant to human users (e.g., depth and width of a seat, height of an overhang).

\begin{figure}[t]
    \centering
    \includegraphics[width=\figurewidth]{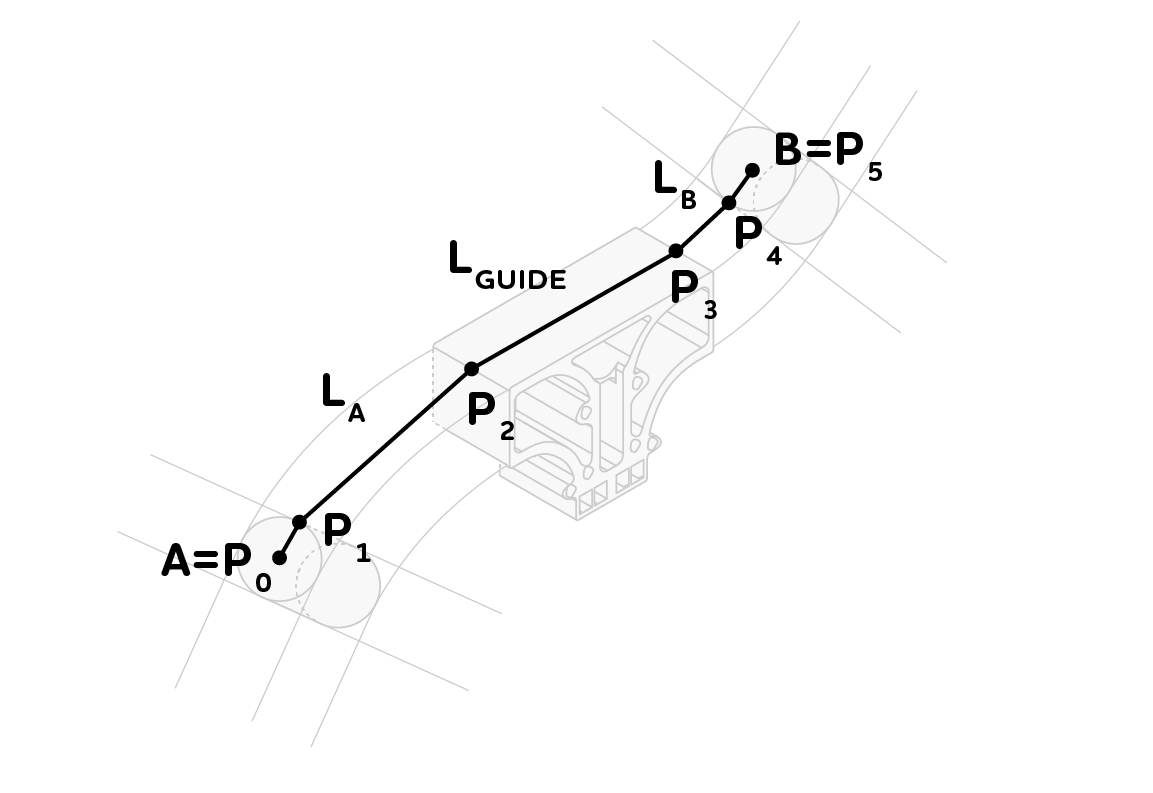}
    \caption{In our simulation, each T-Module is be represented by a polyline consisting of six points. The points correspond to particles in our solver.
    }
    \label{fig:geo}
\end{figure}

\begin{figure}[t]
    \centering
    \includegraphics[width=\figurewidth]{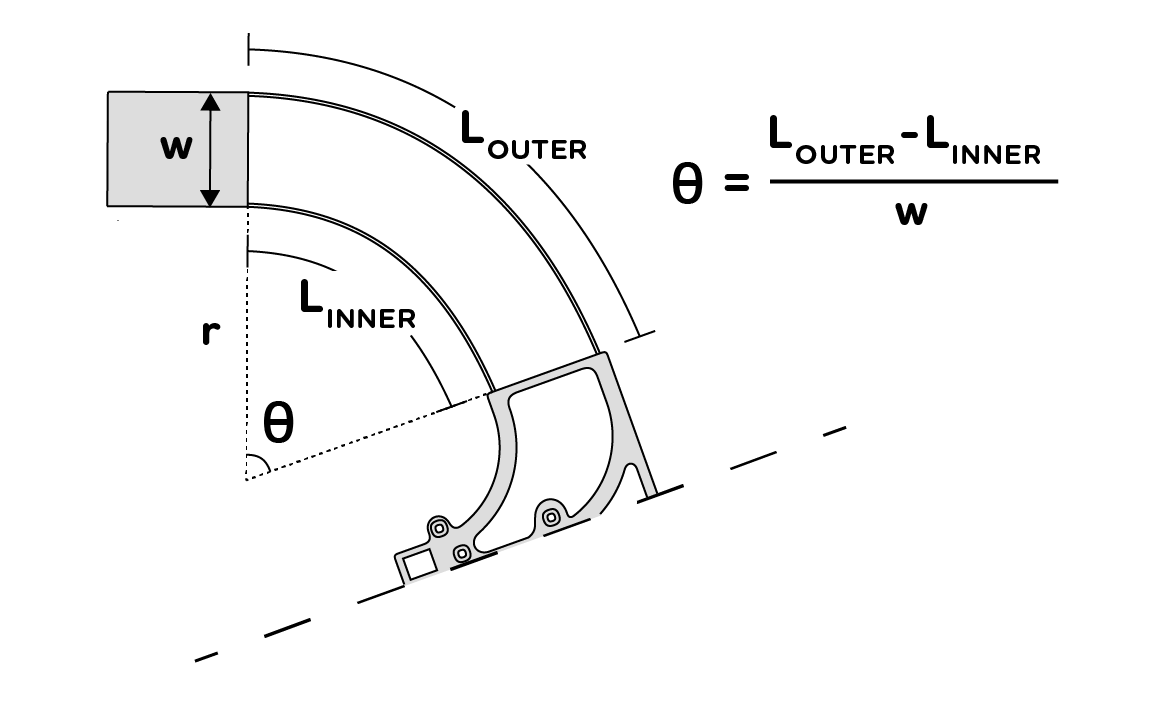}
    \caption{We can compare the measured lengths of the inner and outer bands to detect the direction and amount of curvature.
    }
    \label{fig:geo-curve}
\end{figure}

\subsection{Geometric Representation}

We represent each adjustable-length member with a polyline consisting of both fixed and variable-length segments (\figref{fig:geo}). Six points define this polyline, each mapping to a particle in our simulation.

The two endpoints, $p_{0}$ and $p_{5}$, correspond to the "A" and "B" vertices of the T-module. Two intermediate points, $p_{1}$ and $p_{4}$, are placed on the perimeters of the mechanical vertices (the plastic cylinders where are T-modules joined together). The remaining two points, $p_{2}$ and $p_{3}$, represent the edges of the rigid guide.

In our simulation, particles are shared between mesh members. Specifically, this occurs at the "A" and "B" vertices. For example, $p_{5}$ for one mesh member may be $p_{0}$ for another. Our solver does not distinguish between these shared particles and simply sums all energies acting on each one, resulting in a cohesive mesh structure.

\subsection{Energy Terms}

Our simulation incorporates several types of energy terms that collectively determine the final configuration:

\begin{enumerate}
    \item \emph{Length Energies.} These act like springs between vertices, pushing or pulling particles until the distance between them matches the sensed physical length.
    \item \emph{Coplanarity Energies.} These maintain local smoothness at mechanical vertices by ensuring surrounding points lie approximately in the same plane.
    \item \emph{Bending Energies.} These model the physical stiffness of the FR4 strips, restricting how sharply the mesh can curve.
    \item \emph{Anchor Constraints.} These fix certain points in space, typically along the boundary of the mesh.
    \item \emph{Bias Energies.} These help resolve bistable configurations by pushing vertices in the correct curvature direction.
\end{enumerate}

These energies work in concert, with length energies primarily determining the overall shape, while the others refine local behavior to match physical constraints.

\subsection{Handling Bistability}

When attempting to reconstruct the mesh from length measurements alone, we encounter a fundamental ambiguity: knowing only the distance between endpoints does not tell us whether a curved edge bends inward or outward. This bistability creates a challenge for our reconstruction algorithm.

To address this, we begin with a coarse solver that considers only "length energies" and optimizes just the particles corresponding to the "A" and "B" vertices. These length energies function like springs --- when the distance between $p_{0}$ and $p_{5}$ equals the sensed, real-world distance, the particles exist in a zero-energy state. Otherwise, a correction vector guides the particles toward the correct distance.

However, with length constraints alone, the solver may converge to configurations where areas that should be convex appear concave, or vice versa. When a T-Module is equipped with all four length sensors (both inner and outer bands), we can determine the direction of curvature by comparing these readings (Figure 16). When the outer band is longer than the inner band, we know the module curves away from the outer band.

Once the curvature direction is determined, we apply a constant-magnitude bias force in that direction. This force is "artificial" in that it doesn't model a physical phenomenon but instead helps guide the solver toward the correct solution by breaking the symmetry of bistable configurations. The magnitude of this force remains constant regardless of how much curvature is detected—it merely indicates the correct direction.

% For T-Modules with only two length sensors (lacking inner/outer readings), we do not automatically apply bias forces. In these cases, user guidance may be necessary to resolve ambiguities, especially in underconstrained regions of the mesh.

After the coarse solver has converged, points $p_{1}$ through $p_{4}$ are interpolated between $p_{0}$ and $p_{5}$ as an initial guess for their final positions. We then proceed to reconstruct the curvature by incorporating additional energy terms.

\subsection{Local Coplanarity}

The mechanical vertices of our mesh structure are rigid cylinders, 6~cm in diameter. This physical constraint requires that all points along the circumference of each vertex remain approximately coplanar. We enforce this with a "coplanarity energy" term in our simulation.

For each vertex, we gather the four surrounding particles and fit a plane to all five points (the vertex itself plus its four neighbors). The best-fit plane is calculated using standard least-squares methods. For each of these points, we generate a correction vector perpendicular to the plane, with magnitude equal to the point's distance from the plane.

\begin{figure}[b]
    \centering
    \includegraphics[width=\figurewidth]{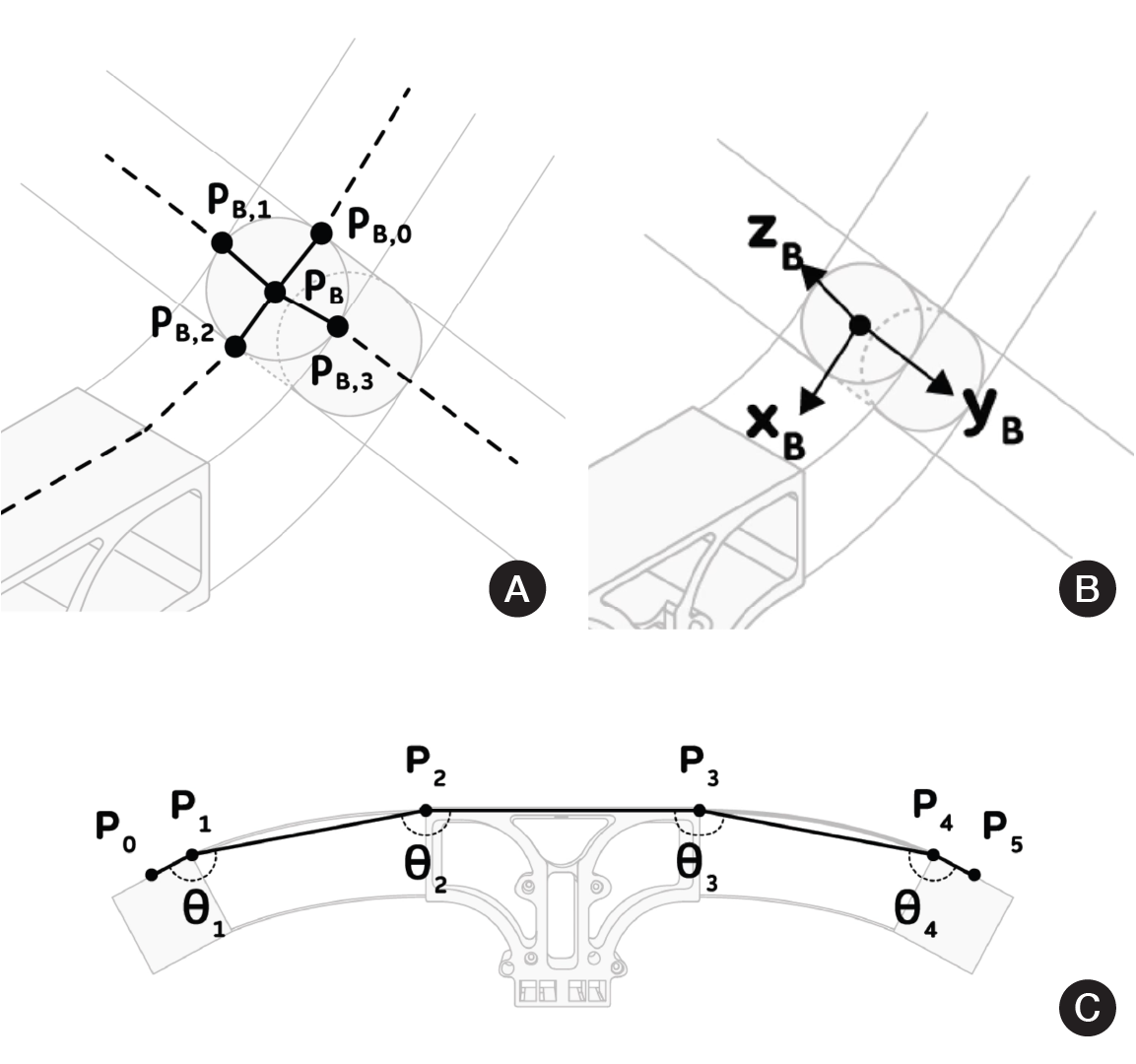}
    \caption{(a) Vertex particles, along with their four immediate neighboring particles, are softly constrained to be coplanar. (b) A reference frame is constructed at each module vertex. (c) A bending energy attempts to minimize the angle between adjacent line segments.
    }
    \label{fig:coplanar}
\end{figure}

When applied to our simulation, this coplanarity constraint has the effect of smoothing the sharp features that may result from the coarse solver. It also effectively "clamps" the ends of the flexible bands, constraining the bending energy that we add in the next section and better reflecting the physical behavior of our T-Modules.

\subsection{Axial Bending}

For every trio of adjacent points in our polyline representation, we calculate the bending stresses using the method described by Adriaenssens and Barnes \cite{adriaenssens2001tensegrity}. This approach models each segment as a rod with internal bending resistance, yielding a shearing transformation that, when applied, drives the member towards its natural curvature.

This method, however, assumes a circular cross-section with a uniform moment of inertia --- an assumption not met by our system. To account for this, we constrain the bending to occur only in the plane where the physical module can actually flex, eliminating forces that would attempt to bend the module in physically impossible ways. Specifically, we project the particle correction vectors onto the plane defined by $p_{0}$, $p_{5}$, and the average of the vertex normals at those points. While this adaptation simplifies the true mechanical behavior, it captures the essential characteristics needed for our reconstruction.

\subsection{Post-Processing for Final Form}

Once the energy minimization simulation converges to a stable configuration that satisfies our measured constraints, we apply several post-processing steps to create a smoother, more usable digital model.

First, we fit natural cubic splines through the polyline representations of each member. This smooths any remaining discontinuities in the simulated mesh while preserving the overall shape and measured dimensions.

After creating splines for all members, we generate Coons patches between adjacent splines. These surface patches interpolate between the boundary curves, creating smooth surface transitions across the entire mesh structure. Finally, the individual Coons patches are joined together to create a unified surface representation of the sculpted form.

This post-processed model can then be directly imported into CAD software for further refinement or manufacturing preparation, as demonstrated in the library bench example in Section 2.
\section{Discussion and Future Work}
Our shape-aware mesh system enables low-fidelity form-finding through tangible interaction, serving as an important link between hands-on prototyping and the digital manufacturing pipeline. While our current implementation demonstrates the potential of this approach, there are several areas where future work could enhance functionality and expand applications.

\subsection{Limitations}
Perhaps the most visible limitation of our system is resolution. While our tool is well-suited for room-scale applications (e.g., benches, alcoves, facades), more detailed forms may demand a larger number of small-scale modules. The current module size creates a tradeoff between workspace coverage and feature resolution --- a challenge when designing objects that contain both broad surfaces and fine details.

Another consideration is the material properties of the mesh. Although the flexible FR4 panels used in our structure are well-suited for modeling curved surfaces, they are subject to small amounts of plastic deformation when held in place over time. This did not noticeably affect the shapes we were able to sculpt during our testing, but it is worth noting as a potential long-term failure point, particularly for installations that might remain in a fixed configuration for extended periods.

Finally, the accuracy of our reconstruction routine is contingent on several assumptions about material behavior. In practice, small inconsistencies in the flexibility of the members, or external forces like gravity, may cause deviations from the predicted shapes.

\subsection{Comparison with Existing Digitization Techniques}

Unlike real-time 3D scanning approaches that capture surface geometry through point clouds, our shape-aware mesh provides direct structural information through edge length measurements. This distinction offers several advantages: our system eliminates occlusion issues common in optical scanning, requires no post-processing to clean up point cloud data, and seamlessly integrates with CAD workflows through a direct data pipeline. Additionally, while scanning approaches passively record geometry, our system actively captures designer intent through direct manipulation of structural elements, creating a more deliberate bridge between physical intuition and digital representation.

Furthermore, traditional form-finding techniques typically operate as one-way processes, either moving from physical models to digital representations (through scanning) or from digital designs to physical models (through fabrication). Our approach establishes a continuous feedback loop where physical manipulations can be immediately visualized digitally, evaluated, and refined. This bidirectional relationship transforms the mesh from a simple digitization tool into an interactive medium that combines the tactile benefits of physical prototyping with the precision and iterative capabilities of digital design.

\subsection{Future Directions in Hardware and Sensing}

To address the resolution limitations, future implementations could incorporate modules of different sizes within the same structure --- allowing for higher-resolution patches alongside coarser features. This multi-scale approach would enable designers to focus on detail where needed, while maintaining manageable complexity in areas requiring less definition.

Alternate materials (with better fatigue resistance) could be explored as well. Composites that combine the resistive properties of our FR4 strips with improved mechanical characteristics might offer an ideal balance between sensing capability and physical behavior.

\hl{Lastly, the neighbor-to-neighbor communication in our mesh could potentially introduce latency for structures that are larger than room-scale (as the number of hops increases). A potential solution is to introduce localized hubs that all share a communication bus, and then   hop from hub to hub instead of module to module.}

\subsection{Topologies and Configurations}
Although our current implementation requires users to predefine the mesh topology, our system is also capable of supporting automatic topology discovery. By leveraging the communication capabilities of the modules, each unit is able to identify its connected neighbors by exchanging messages and recording connection points. Future implementations can leverage this existing feature --- dynamically constructing the mesh topology without manual input, and streamlining the user experience.

\begin{figure}[t]
  \centering
  \includegraphics[width=\figurewidth]{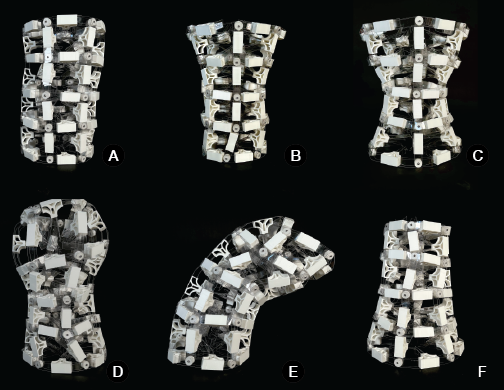}
  \caption{Our modules are capable of starting from geometries other than flat planes. Here we explore various forms based on a cylinder. }
  \Description{}
\end{figure}

% \begin{figure}[h] \centering \includegraphics[width=\textwidth]{figures/cylinder.png} \caption{Our modules are capable of starting from geometries other than flat planes. Here we explore various forms based on a cylinder. } \label{fig:cylinder} \end{figure}

Beyond flat meshes, our system can support various starting geometries, such as the cylindrical configurations in Figure 18. Future work could explore more complex initial topologies: spheres, toroidal surfaces, or other structures -- expanding the design space for architectural applications.

\subsection{Bidirectional Information Flow}

In this work, we have focused primarily on the flow of information from the physical to the digital. Moving forward, it may also be valuable to explore the reverse: enabling mesh structures with actuated, length-changing edges. Such a system would allow a seasoned CAD user to first create a form in software, and then experience it in physical space—perhaps making hands-on refinements when design issues become apparent. Since each module is already equipped with absolute length sensors, closed-loop control of this shape-changing mesh may be feasible without significant hardware modifications.

A system capable of physical-to-digital and digital-to-physical transformations could serve as a platform for exploring the interplay between computational optimization and human intuition. Algorithms could suggest modifications to human-created forms to improve structural performance or material efficiency, then physically manifest these changes for designer evaluation.

Such a system could also allow for collaborative design processes within geographically distributed teams. Architects in different locations could manipulate the same structure --- changes made to a physical mesh in one location could be transmitted digitally and reproduced elsewhere (e.g. on an actuated mesh, or in virtual reality).

\subsection{Applications Beyond Architecture}

\begin{figure}[t]
\centering
\includegraphics[width=\figurewidth]{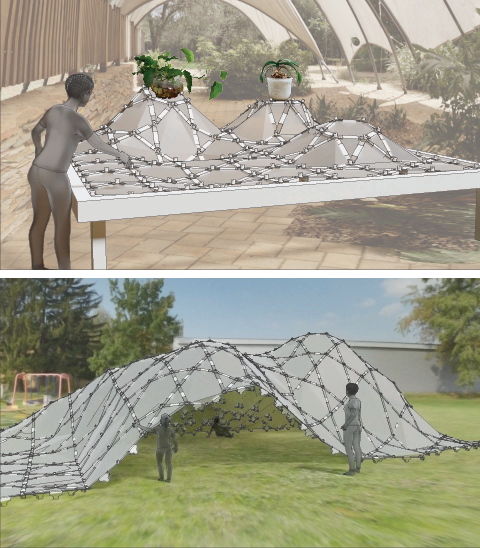}
\caption{Envisioned uses. Top: An exhibit designer working with difficult-to-digitize items, such as plants, can arrange them directly on a physical display table. Bottom: Children can help design a play space without needing to understand scale models or virtual renderings.
}
\label{fig}
\end{figure}

The system also shows promise for accessibility and inclusive design. For individuals who may struggle with traditional CAD interfaces, a tangible modeling system provides an alternative pathway to spatial design. This could be particularly valuable in participatory design sessions involving diverse stakeholders with varying technical backgrounds, as illustrated in \figref{fig} (bottom), where children can directly engage in designing play spaces without needing to understand scale models or interpret virtual renderings.

In museum and exhibition design, our system offers unique advantages for working with difficult-to-digitize objects. As shown in \figref{fig} (top), exhibition designers can arrange physical items like plants directly on a shape-aware surface, sculpting display fixtures around them in real time. The resulting configurations can be immediately captured in digital form, bypassing the typical challenges of 3D scanning organic objects with complex geometries.

\section{Conclusion} In this paper, we introduced an interactive, large-scale sculptable mesh structure that bridges the gap between physical form-finding and digital design processes in architecture. Our system consists of adjustable-length members equipped with onboard sensors that detect changes in edge lengths, allowing for real-time digital reconstruction of the sculpted geometry. This enables architects, designers, and even community participants to physically manipulate mesh structures and immediately see the impact of their changes in a digital model.

Through a fictional participatory design scenario involving a library renovation, we demonstrated how users can intuitively shape the mesh to create various architectural elements, such as alcoves, shelves, and seating areas. The immediate physical feedback provided by this low-fidelity prototyping allows users to test and refine their ideas on location. Physical adjustments made to the sculpted structure can be captured, reconstructed by our energy-minimization routine, and subsequently imported into CAD software for post-processing and fabrication. Through the type of enhanced user engagement that our system facilitates, we hope to further enrich the built environment.

\begin{acks}
The work in this paper was
supported in part by the National Science Foundation under Award No.
2420434.
\end{acks}

% \begin{acks}
%     We thank Hatice Gokcen Guner for help with setting up the gauge and application prototyping, Dinesh K. Patel and Hugo de Souza Oliveira for their expertise in SMA, Shuhong Wang for skillful photo shooting and editing, and Humphery Yang for valuable suggestions on paper writing. Special thanks to Xiaolei Lu's unwavering support.
% \end{acks}

\bibliographystyle{ACM-Reference-Format}
\bibliography{references}

@inproceedings{bacher2016defsense,
  title={Defsense: Computational design of customized deformable input devices},
  author={B{\"a}cher, Moritz and Hepp, Benjamin and Pece, Fabrizio and Kry, Paul G and Bickel, Bernd and Thomaszewski, Bernhard and Hilliges, Otmar},
  booktitle={Proceedings of the 2016 CHI Conference on Human Factors in Computing Systems},
  pages={3806--3816},
  year={2016}
}

@inproceedings{balakrishnan1999exploring,
  title={Exploring interactive curve and surface manipulation using a bend and twist sensitive input strip},
  author={Balakrishnan, Ravin and Fitzmaurice, George and Kurtenbach, Gordon and Singh, Karan},
  booktitle={Proceedings of the 1999 symposium on Interactive 3D graphics},
  pages={111--118},
  year={1999}
}

@inproceedings{chien2015flexibend,
  title={Flexibend: Enabling interactivity of multi-part, deformable fabrications using single shape-sensing strip},
  author={Chien, Chin-yu and Liang, Rong-Hao and Lin, Long-Fei and Chan, Liwei and Chen, Bing-Yu},
  booktitle={Proceedings of the 28th Annual ACM Symposium on User Interface Software \& Technology},
  pages={659--663},
  year={2015}
}

@inproceedings{wessely2018shape,
  title={Shape-aware material: Interactive fabrication with shapeme},
  author={Wessely, Michael and Tsandilas, Theophanis and Mackay, Wendy E},
  booktitle={Proceedings of the 31st Annual ACM Symposium on User Interface Software and Technology},
  pages={127--139},
  year={2018}
}

@inproceedings{sheng2006interface,
  title={An interface for virtual 3D sculpting via physical proxy.},
  author={Sheng, Jia and Balakrishnan, Ravin and Singh, Karan},
  booktitle={GRAPHITE},
  volume={6},
  pages={213--220},
  year={2006}
}

@article{adriaenssens2001tensegrity,
  title={Tensegrity spline beam and grid shell structures},
  author={Adriaenssens, SMLa and Barnes, Michael R},
  journal={Engineering structures},
  volume={23},
  number={1},
  pages={29--36},
  year={2001},
  publisher={Elsevier}
}

@article{jones2020characterising,
  title={Characterising the Digital Twin: A systematic literature review},
  author={Jones, David and Snider, Chris and Nassehi, Aydin and Yon, Jason and Hicks, Ben},
  journal={CIRP journal of manufacturing science and technology},
  volume={29},
  pages={36--52},
  year={2020},
  publisher={Elsevier}
}

@inproceedings{weller2008posey,
  title={Posey: instrumenting a poseable hub and strut construction toy},
  author={Weller, Michael Philetus and Do, Ellen Yi-Luen and Gross, Mark D},
  booktitle={Proceedings of the 2nd international conference on Tangible and embedded interaction},
  pages={39--46},
  year={2008}
}

@inproceedings{leen2017strutmodeling,
  title={StrutModeling: A low-fidelity construction kit to iteratively model, test, and adapt 3D objects},
  author={Leen, Danny and Ramakers, Raf and Luyten, Kris},
  booktitle={Proceedings of the 30th Annual ACM Symposium on User Interface Software and Technology},
  pages={471--479},
  year={2017}
}

@inproceedings{glaessgen2012digital,
  title={The digital twin paradigm for future NASA and US Air Force vehicles},
  author={Glaessgen, Edward and Stargel, David},
  booktitle={53rd AIAA/ASME/ASCE/AHS/ASC structures, structural dynamics and materials conference 20th AIAA/ASME/AHS adaptive structures conference 14th AIAA},
  pages={1818},
  year={2012}
}

@inproceedings{feick2020tangi,
  title={Tangi: Tangible proxies for embodied object exploration and manipulation in virtual reality},
  author={Feick, Martin and Bateman, Scott and Tang, Anthony and Miede, Andr{\'e} and Marquardt, Nicolai},
  booktitle={2020 IEEE international symposium on mixed and augmented reality (ISMAR)},
  pages={195--206},
  year={2020},
  organization={IEEE}
}

@inproceedings{piya2014proto,
  title={Proto-tai: Quick design prototyping using tangible assisted interfaces},
  author={Piya, Cecil and Ramani, Karthik},
  booktitle={International Design Engineering Technical Conferences and Computers and Information in Engineering Conference},
  volume={46285},
  pages={V01AT02A097},
  year={2014},
  organization={American Society of Mechanical Engineers}
}

@incollection{singh2006industrial,
  title={Industrial motivation for interactive shape modeling: a case study in conceptual automotive design},
  author={Singh, Karan},
  booktitle={Acm siggraph 2006 courses},
  pages={3--9},
  year={2006}
}

@article{piker2013kangaroo,
  title={Kangaroo: form finding with computational physics},
  author={Piker, Daniel},
  journal={Architectural Design},
  volume={83},
  number={2},
  pages={136--137},
  year={2013},
  publisher={Wiley Online Library}
}

@inproceedings{gulay2021understanding,
  title={Understanding the Role of Physical and Digital Techniques in the Initial Design Processes of Architecture},
  author={Gulay, Emrecan and Lucero, Andr{\'e}s},
  booktitle={IFIP Conference on Human-Computer Interaction},
  pages={312--329},
  year={2021},
  organization={Springer}
}

@inproceedings{willis2010interactive,
  title={Interactive fabrication: new interfaces for digital fabrication},
  author={Willis, Karl DD and Xu, Cheng and Wu, Kuan-Ju and Levin, Golan and Gross, Mark D},
  booktitle={Proceedings of the fifth international conference on Tangible, embedded, and embodied interaction},
  pages={69--72},
  year={2010}
}

@inproceedings{baumgart1975polyhedron,
  title={A polyhedron representation for computer vision},
  author={Baumgart, Bruce G},
  booktitle={Proceedings of the May 19-22, 1975, national computer conference and exposition},
  pages={589--596},
  year={1975}
}

@inproceedings{agrawal2015protopiper,
  title={Protopiper: Physically sketching room-sized objects at actual scale},
  author={Agrawal, Harshit and Umapathi, Udayan and Kovacs, Robert and Frohnhofen, Johannes and Chen, Hsiang-Ting and Mueller, Stefanie and Baudisch, Patrick},
  booktitle={Proceedings of the 28th Annual ACM Symposium on User Interface Software \& Technology},
  pages={427--436},
  year={2015}
}

@inproceedings{tahouni2020nurbsforms,
  title={NURBSforms: a modular shape-changing interface for prototyping curved surfaces},
  author={Tahouni, Yasaman and Qamar, Isabel PS and Mueller, Stefanie},
  booktitle={Proceedings of the Fourteenth International Conference on Tangible, Embedded, and Embodied Interaction},
  pages={403--409},
  year={2020}
}

@article{khajavi2019digital,
  title={Digital twin: vision, benefits, boundaries, and creation for buildings},
  author={Khajavi, Siavash H and Motlagh, Naser Hossein and Jaribion, Alireza and Werner, Liss C and Holmstr{\"o}m, Jan},
  journal={IEEE access},
  volume={7},
  pages={147406--147419},
  year={2019},
  publisher={IEEE}
}

@article{stappers2009designing,
  title={Designing for other people’s strengths and motivations: Three cases using context, visions, and experiential prototypes},
  author={Stappers, Pieter Jan and van Rijn, Helma and Kistemaker, SC and Hennink, AE and Visser, F Sleeswijk},
  journal={Advanced engineering informatics},
  volume={23},
  number={2},
  pages={174--183},
  year={2009},
  publisher={Elsevier}
}

@article{brandt2007tangible,
  title={How tangible mock-ups support design collaboration},
  author={Brandt, Eva},
  journal={Knowledge, Technology \& Policy},
  volume={20},
  number={3},
  pages={179--192},
  year={2007},
  publisher={Springer}
}

@article{sanders2014probes,
  title={Probes, toolkits and prototypes: three approaches to making in codesigning},
  author={Sanders, Elizabeth B-N and Stappers, Pieter Jan},
  journal={CoDesign},
  volume={10},
  number={1},
  pages={5--14},
  year={2014},
  publisher={Taylor \& Francis}
}

@article{luck2018participatory,
  title={Participatory design in architectural practice: Changing practices in future making in uncertain times},
  author={Luck, Rachael},
  journal={Design Studies},
  volume={59},
  pages={139--157},
  year={2018},
  publisher={Elsevier}
}

@article{sanoff2002schools,
  title={Schools Designed with Community Participation.},
  author={Sanoff, Henry},
  year={2002},
  publisher={ERIC}
}

@article{sanoff2005community,
  title={Community participation in riverfront development},
  author={Sanoff, Henry},
  journal={CoDesign},
  volume={1},
  number={1},
  pages={61--78},
  year={2005},
  publisher={Taylor \& Francis}
}

@book{sanoff1999community,
  title={Community participation methods in design and planning},
  author={Sanoff, Henry},
  year={1999},
  publisher={John Wiley \& Sons}
}

@article{abrams2003byker,
  title={Byker revisited},
  author={Abrams, Robin},
  journal={Built Environment (1978-)},
  pages={117--131},
  year={2003},
  publisher={JSTOR}
}

@inproceedings{buchenau2000experience,
  title={Experience prototyping},
  author={Buchenau, Marion and Suri, Jane Fulton},
  booktitle={Proceedings of the 3rd conference on Designing interactive systems: processes, practices, methods, and techniques},
  pages={424--433},
  year={2000}
}

@article{lim2008anatomy,
  title={The anatomy of prototypes: Prototypes as filters, prototypes as manifestations of design ideas},
  author={Lim, Youn-Kyung and Stolterman, Erik and Tenenberg, Josh},
  journal={ACM Transactions on Computer-Human Interaction (TOCHI)},
  volume={15},
  number={2},
  pages={1--27},
  year={2008},
  publisher={ACM New York, NY, USA}
}

@book{buxton2010sketching,
  title={Sketching user experiences: getting the design right and the right design},
  author={Buxton, Bill},
  year={2010},
  publisher={Morgan kaufmann}
}

@article{evans2005rapid,
  title={Rapid prototyping and industrial design practice: can haptic feedback modelling provide the missing tactile link?},
  author={Evans, Mark A},
  journal={Rapid Prototyping Journal},
  volume={11},
  number={3},
  pages={153--159},
  year={2005},
  publisher={Emerald Group Publishing Limited}
}

@book{binder2011design,
  title={Design things},
  author={Binder, Thomas and De Michelis, Giorgio and Ehn, Pelle and Jacucci, Giulio and Linde, Per},
  year={2011},
  publisher={MIT press}
}

@article{schon1992designing,
  title={Designing as reflective conversation with the materials of a design situation},
  author={Schon, Donald A},
  journal={Research in engineering design},
  volume={3},
  number={3},
  pages={131--147},
  year={1992},
  publisher={Springer}
}

@book{pallasmaa_eyes_2012,
  title = {The Eyes of the Skin: Architecture and the Senses},
  shorttitle = {The Eyes of the Skin},
  author = {Pallasmaa, Juhani},
  date = {2012},
  publisher = {{Wiley-Academy ; John Wiley \& Sons}},
  location = {{Chichester : Hoboken, NJ}},
  isbn = {978-0-470-01579-7 978-0-470-01578-0},
  langid = {english},
  pagetotal = {128}
}

@article{greenberg_value_2017,
  title={The value of the full-scale prototype in architectural education.},
  author={Greenberg, Evan},
  journal={Charrette},
  volume={4},
  number={1},
  pages={27--39},
  year={2017},
  publisher={association of architectural educators (aae)}
}

@inproceedings{gonzalez2023constraint,
  title={Constraint-Driven Robotic Surfaces, At Human-Scale},
  author={Gonzalez, Jesse T and Prashant, Sonia and Tayal, Sapna and Kedia, Juhi and Ion, Alexandra and Hudson, Scott E},
  booktitle={Proceedings of the 36th Annual ACM Symposium on User Interface Software and Technology},
  pages={1--12},
  year={2023}
}

@inproceedings{smith2025voxel,
  title={Voxel Invention Kit: Reconfigurable Building Blocks for Prototyping Interactive Electronic Structures},
  author={Smith, Miana and Forman, Jack and Abdel-Rahman, Amira and Wang, Sophia and Gershenfeld, Neil},
  booktitle={Proceedings of the 2025 CHI Conference on Human Factors in Computing Systems},
  pages={1--15},
  year={2025}
}

@inproceedings{yu2022strawctures,
  title={Strawctures: A modular electronic construction kit for human-scale interactive structures},
  author={Yu, Jin and Sakhardande, Prabodh and Parmar, Ruchita and Oh, HyunJoo},
  booktitle={Proceedings of the Sixteenth International Conference on Tangible, Embedded, and Embodied Interaction},
  pages={1--10},
  year={2022}
}

@inproceedings{rambold2023airtied,
  title={AirTied: Automatic Personal Fabrication of Truss Structures},
  author={Rambold, Lukas and Kovacs, Robert and Lempert, Conrad and Abdullah, Muhammad and Lendowski, Helena and Fritzsche, Lukas and Taraz, Martin and Baudisch, Patrick},
  booktitle={Proceedings of the 36th Annual ACM Symposium on User Interface Software and Technology},
  pages={1--10},
  year={2023}
}

@inproceedings{je2021elevate,
  title={Elevate: A walkable pin-array for large shape-changing terrains},
  author={Je, Seungwoo and Lim, Hyunseung and Moon, Kongpyung and Teng, Shan-Yuan and Brooks, Jas and Lopes, Pedro and Bianchi, Andrea},
  booktitle={Proceedings of the 2021 CHI Conference on human Factors in Computing Systems},
  pages={1--11},
  year={2021}
}

@inproceedings{suzuki2020roomshift,
  title={Roomshift: Room-scale dynamic haptics for vr with furniture-moving swarm robots},
  author={Suzuki, Ryo and Hedayati, Hooman and Zheng, Clement and Bohn, James L and Szafir, Daniel and Do, Ellen Yi-Luen and Gross, Mark D and Leithinger, Daniel},
  booktitle={Proceedings of the 2020 CHI conference on human factors in computing systems},
  pages={1--11},
  year={2020}
}

@article{swaminathan2019input,
  title={Input, output and construction methods for custom fabrication of room-scale deployable pneumatic structures},
  author={Swaminathan, Saiganesh and Rivera, Michael and Kang, Runchang and Luo, Zheng and Ozutemiz, Kadri Bugra and Hudson, Scott E},
  journal={Proceedings of the ACM on Interactive, Mobile, Wearable and Ubiquitous Technologies},
  volume={3},
  number={2},
  pages={1--17},
  year={2019},
  publisher={ACM New York, NY, USA}
}

@inproceedings{grossman2003interface,
  title={An interface for creating and manipulating curves using a high degree-of-freedom curve input device},
  author={Grossman, Tovi and Balakrishnan, Ravin and Singh, Karan},
  booktitle={Proceedings of the SIGCHI conference on Human factors in computing systems},
  pages={185--192},
  year={2003}
}

@inproceedings{shahmiri2020sharc,
  title={Sharc: A geometric technique for multi-bend/shape sensing},
  author={Shahmiri, Fereshteh and Dietz, Paul H},
  booktitle={Proceedings of the 2020 CHI Conference on Human Factors in Computing Systems},
  pages={1--12},
  year={2020}
}

@inproceedings{lee2000handscape,
  title={HandSCAPE: a vectorizing tape measure for on-site measuring applications},
  author={Lee, Jay and Su, Victor and Ren, Sandia and Ishii, Hiroshi},
  booktitle={Proceedings of the SIGCHI conference on Human Factors in Computing Systems},
  pages={137--144},
  year={2000}
}

@inproceedings{dementyev2015sensortape,
  title={Sensortape: Modular and programmable 3d-aware dense sensor network on a tape},
  author={Dementyev, Artem and Kao, Hsin-Liu and Paradiso, Joseph A},
  booktitle={Proceedings of the 28th Annual ACM Symposium on User Interface Software \& Technology},
  pages={649--658},
  year={2015}
}

@inproceedings{weichel2015spata,
  title={SPATA: Spatio-tangible tools for fabrication-aware design},
  author={Weichel, Christian and Alexander, Jason and Karnik, Abhijit and Gellersen, Hans},
  booktitle={Proceedings of the ninth international conference on tangible, embedded, and embodied interaction},
  pages={189--196},
  year={2015}
}

\appendix

\end{document}